\documentclass[12pt,a4paper,twoside]{article}
\title{} 
\author{} 
\date{} 

\usepackage{hyperref}
\usepackage{amsmath,amssymb,latexsym,amsfonts}
\usepackage{bm}
\usepackage{eurosym}
\usepackage{times} 
\usepackage{rotating}
\usepackage[dvipsnames]{xcolor}
\usepackage{colortbl}
\usepackage{multirow}
\usepackage{graphicx}
\usepackage{caption}
\usepackage{subcaption}
\usepackage{geometry}
\usepackage{enumitem}
\usepackage{caption}
\usepackage{subcaption}
\usepackage{marginnote}
\usepackage{booktabs}
\usepackage{gensymb}
\usepackage{soul}
\usepackage{color}
\usepackage{tikz}
\usepackage{float}
\usepackage{multicol}
\usepackage{natbib}
\usepackage{pdflscape}
\usepackage[parfill]{parskip}
\usepackage{geometry}
\usetikzlibrary{automata}

\newlength{\Oldarrayrulewidth}

\usepackage{natbib} 
\bibpunct{(}{)}{,}{}{}{,} 

\usepackage{url} 

\usepackage{graphicx} 
\usepackage{float}
\restylefloat{figure} \restylefloat{table}
\usepackage[bf,tiny]{titlesec}
\titleformat{\section}{\bfseries}{\thesection.}{1ex}{}
\titlespacing*{\section}{0pt}{8ex}{3ex}
\titleformat{\subsection}{\bfseries}{\thesubsection.}{1ex}{}
\titlespacing*{\subsection}{0pt}{6ex}{2ex}
\titleformat{\subsubsection}{\scshape}{\thesubsubsection.}{1ex}{}
\titlespacing*{\subsubsection}{0pt}{6ex}{2ex}
\newtheorem{theorem}{Theorem}

\newtheorem{proposition}[theorem]{Proposition}

\newtheorem{remark}{Remark}
\newenvironment{proof}[1][Proof]{\noindent \textbf{#1.} }{\hfill
\rule{0.5em}{0.5em}}

\usepackage{fancyhdr}
\renewcommand\appendix{\par
\setcounter{section}{0}
\setcounter{subsection}{0}
\setcounter{table}{0}
\renewcommand\thesection{Appendix \Alph{section}}
\renewcommand\thetable{\Alph{section}\arabic{table}}
}
\begin{document}
\setcounter{page}{1}
\setcounter{equation}{0}
\setcounter{section}{0}
\setcounter{table}{0}
\setcounter{figure}{0}
\setcounter{footnote}{0}
\thispagestyle{plain}

\fancyhead[LE,RO]{\thepage} \fancyhead[LO]{\sl Measuring systemic risk and contagion in the European financial network}
\vspace*{1.7cm}
\begin{center}
\normalsize 
\textbf{Measuring systemic risk and contagion in the European financial network}\\
\author{Laleh}
\bigskip
\noindent \textit{ Laleh Tafakori{$^\dagger$} \footnote{{\textit {{$^\dagger$}Address for correspondence:}} Department of Mathematical Sciences, School of Science, RMIT University, Melbourne, Australia, Email:  laleh.tafakori@rmit.edu.au.}},
\noindent \textit{ Armin Pourkhanali \footnote{Centre for Global Business, Monash University, Australia, Email: armin.pourkhanali@monash.edu.}},
\noindent \textit{ Riccardo Rastelli \footnote{ School of Mathematics and Statistics, University College Dublin, Ireland, Email: riccardo.rastelli@ucd.ie}}
\bigskip
\\
\today
\end{center}
\mbox{ }\\
\begin{abstract}
\noindent
 This paper introduces a novel framework to study default dependence and systemic risk in a financial network that evolves over time. We analyse several indicators of risk, and develop a new latent space model to assess the health of key European banks before, during, and after the recent financial crises. First, we adopt the measure of CoRisk to determine the impact of such crises on the financial network. 
Then, we use minimum spanning trees to analyse the correlation structure and the centrality of the various banks.
Finally, we propose a new statistical model that permits a latent space visualisation of the financial system.\footnote{A package in R  is developed to implement Latent space models for inference and visualisation of the financial system. This package contains several functions which use different statistical approaches. } This provides a clear and interpretable model-based summary of the interaction data, and it gives a new perspective on the topology structure of the network.
Crucially, the methodology provides a new approach to assess and understand the systemic risk associated with a financial system, and to study how debt may spread between institutions. Our dynamic framework provides an interpretable map that illustrates the default dependencies between institutions, highlighting the possible patterns of contagion and the institutions that may pose systemic threats.

\bigskip

\noindent {\color{black}{\textit Keywords:}  Finance; Default probability; Financial contagion; Latent space models; Financial networks}

\end{abstract}

\bigskip\bigskip\bigskip
\normalsize
\section{Introduction}\label{Intro}
Over the last decade, the global financial system has endured several major crises, most notably the Global Financial Crisis of 2008 and the European Sovereign Debt Crisis in 2010. These events raised serious concerns regarding the resilience of the financial system against the contagion of debt during periods of economic turmoil. 
The collapse of Lehman Brothers on September 15, 2008, was the largest bankruptcy filing in U.S. history. This financial shock escalated into a global crisis which clearly demonstrated the frailty of the financial ecosystem, and the ineffectiveness of the current regulations.
Subsequently, many repercussions were felt throughout the world by the participants of the global financial market, resulting in an excess of cross-border and cross-entity interdependencies (see \cite{de2012international} and \cite{acharya2014pyrrhic}).
By the end of September 2008, the shock had rapidly spread across Europe, where the Euro area governments rescued the Belgian-French bank Dexia. 
Many authors agree on the fact that the increased interdependecies between the institutions has been playing a crucial role in the spread of contagion, and in forcing hasty responses to the shocks observed in the system (\cite{aiyar2012financial},\cite{ acemoglu2015systemic}). 
\\
More recently, the research on financial networks and systemic risk has developed in many different directions (\cite{bartram2007estimating}, \cite{schweitzer2009economic}, \cite{engle2014systemic}).
A special emphasis has been dedicated to the understanding of how the topology of the system may impact the potential spread of contagion and systemic risk (\cite{elliott2014financial}, \cite{acemoglu2015systemic}). 
This research question is of particular importance for both regulators and financial institutions, since these should both be able to clearly identify the primary sources of systemic risk, \cite{acharya2012capital}, and thus how the risk may be minimised.
\\
In practice, we are interested in extracting from the financial system a collection of summaries that can permit an assessment of the risk.
In recent years, a number of different measures have been introduced or adapted for this purpose (\cite{Paolo2016}, \cite{battiston2012debtrank}, \cite{friel2016interlocking}, \cite{hledik2018dynamic}).
In this paper, we study a dataset of default probabilities for a set of key European banks from 2005 to 2016, and we use a number of measures to assess the risk associated with each of the institutions.
First, we use a collection of descriptive statistics and other known indicators that have been introduced in financial networks analyses.
Then, we propose a new statistical model that aims at giving a clear and interpretable visualisation of the system by embedding it in a low dimensional space.
\\
At the foundations of our contribution we have our constructed measures of dependencies between any given institutions.
In general, these may be computed using a number of approaches.
One type of approach relies on Pearson's correlation index (see, e.g., \cite{chi2010network}, \cite{wang2015correlation} and \cite{birch2016analysis}).
A second approach uses instead the partial correlation index, which measures and filters a network using partial correlation coefficients between banks. The correlation between two financial agents is frequently influenced by other financial agents, that is, two interacting financial agents may also have correlation with other financial agents (see, e.g., \cite{mantegna1999introduction}). For instance, the US stock market or the European stock markets can affect on the Hong Kong and Chinese stock markets. Therefore, one can get the pure correlation between the Chinese and Hong Kong stock markets by removing any effects of the US and European stock markets. The partial correlation coefficient quantifies the pure correlation between any two financial agents by measuring the relation between them and deducting the impact of any other financial agents. The approaches that make use of this indicator include  \cite{kenett2015partial}, \cite{wang2016tail} \cite{giner2018correlation} and \cite{wen2019tail}.
A third type of approach relies on other correlation-based network methods to construct a network using other similarity measures of correlation (\cite{brida2010dynamics}, \cite{matesanz2014network}). For instance, \cite{brida2010dynamics} introduced the tool of symbolic time series analysis to obtain a metric distance between two different stocks; then, the authors used a minimum spanning tree for investigating the correlation structure of the 30 largest North American companies. Furthermore, \cite{matesanz2014network} studied the non-linear co-movements of foreign exchange markets during the Asian currency crisis by combing the minimum spanning tree and phase synchronisation coefficients. The minimum spanning tree is most frequently used since it is a robust, simple and clear tool to visualise the links. 
\\
In this paper, we transform the partial correlations to obtain the CoRisk measure (\cite{Paolo2016}), which quantifies the difference between the unconditional and the conditional probability of default for an institution. This measure effectively captures how much of the risk associated with an institution is due to the risk of its neighbours. We look at the pairwise CoRisk values between the European banks from 2005 to 2016, hence studying the evolution of this index over time. Then, we propose a different visualisation of the network using minimum spanning trees throughout the same time period, and highlight the importance of nodes using network centrality measures.
\\
Furthermore, we develop a new statistical model that may be used to analyse the nodal attributes of a binary or nonnegatively weighted network.
The framework that we consider is inspired by the literature on Latent Position Models (LPMs, \cite{hoff2002latent}), which has been greatly developed in the last two decades \cite{handcock2007model}, \cite{rastelli2016properties}, \cite{durante2017nonparametric}.
LPMs postulate that the nodes of a graph are embedded in a latent space (usually $\mathbb{R}^2$), and that a connection between any two nodes becomes more likely if the nodes are close to each other, and less likely if they are far apart.
For example, in the context of the recent financial crises, \cite{friel2016interlocking} introduce a type of LPM to study the boards' compositions for companies quoted on the Irish Stock Exchange. 
Their approach provides an easy-to-interpret latent space representation of the Irish financial market, and it leads to the introduction of a new potential measure of financial instability.
Here, our goal is to offer a new latent space perspective on our data, to ultimately assess the health of European banks during the recent crises.
\\
The rest of this paper is structured as follows. Section \ref{sec:methodology} describes the proposed methodologies, inference and model interpretation. Section \ref{sec:results} shows the main empirical results for latent position model and analysing the minimum spanning tree networks by using the default probabilities and CoRisk measure. Finally, we provide conclusions with some discussions in Section \ref{sec:conclusions}.

\section{Methodology}\label{sec:methodology}
\subsection{Default probabilities and CoRisk}
CoRisk is a measure introduced by  \cite{Paolo2016} to determine the variation in the probability of default due to contagion effects. The CoRisk measure consists of two components. Firstly, the additional risk taken on by a financial institution due to its connections with other financial institutions is known as CoRisk$_{in}$. Secondly, the risk caused by the financial institution to other financial institutions that are connected to is known as CoRisk$_{out}$. \\ \\
The values for CoRisk$_{in}$ and CoRisk$_{out}$ (for Bank $j$) can be obtained using the following equations,

\begin{equation}
{CoRisk_{in}}^j = 1 - \prod_{i \in ne(j)} (1 - PD^i)^{\rho_{ij | S}},
\end{equation}
\begin{equation}
{CoRisk_{out}}^j = 1 - (1 - \prod_{i \in ne(j)} (1 - PD^j)^{\rho_{ij | S}}) ,
\end{equation}

where $ne(j)$ represents the neighbours of bank $j$, $PD^i$ is the probability of default for bank $i$ and $\rho_{ij | S}$ is the partial correlation value of banks $i$ and $j$ given $S$ where $S$ is a set of all other banks. 

In order to determine which banks are interconnected in a financial network, a partial correlation matrix is calculated and the partial correlation values are tested for significance. Banks that have a significant partial correlation value with each other are connected by edges in the financial network. We use partial correlation instead of Pearson's correlation as it provides a more accurate view of the correlation between two banks without the influence of external banks.

Moreover, we are interested in pairwise  CoRisk  values between any 2 banks to see which connections between banks are the most significant. The CoRisk value between two banks, bank $i$ and bank $j$, is defined as follows:
\begin{equation}
CoRisk_{ij} = 1 - (1 - APD^{i})^{\rho_{ij}},
\end{equation}
where $APD^{i}$ is the average probability of default of bank $i$ across a specified time period, and $\rho_{ij}$ is the partial correlation value between bank $i$ and bank $j$ over the same time period. Intuitively, this can be interpreted as the effect that the connection of $i$ and $j$ has on the probability of default of bank $j$. 

The CoRisk$_{ij}$ measure has some useful properties. Firstly,  CoRisk$_{ij} < 0$ if and only if two banks have a negative partial correlation value. It takes a value between $0$ and $1$, if the two banks have a positive partial correlation value instead. Secondly, the measure is not symmetric, so  CoRisk$_{ij} \neq$ CoRisk$_{ji}$ unless both banks have the same average default probability. Also, summing up the pairwise CoRisk values from bank $j$ to other banks and from other banks to bank $j$ gives us the aggregate measure of  CoRisk$_{out}$ and  CoRisk$_{in}$ values of the bank respectively. 

\subsection{Construction and interpretation of the adjacency matrices}\label{sec:X}
The observed data consists of the values $\textbf{Y} = \left\{ y_i^{(t)}\in\mathbb{R} \middle \vert i=1,\dots,N; t=1,\dots,T \right\}$, which denote the log default probabilities, and the edges $\textbf{X} = \left\{ x_{ij}^{(t)} \in \{0,1\} \middle\vert i,j = 1, \dots, N; i \neq j \right\}$, which indicate whether two institutions are highly correlated or not.
Here, $N$ represents the number of financial institutions, and $t=1,\dots, T$ determines the time period considered.
The graphs are undirected by construction, that is the edges do not have an orientation and $x_{ij}^{(t)} = x_{ji}^{(t)}$.

For each $t=1,\dots,T$, the matrix $\textbf{X}^{(t)}$ can be seen as the binary {\color{black}adjacency } matrix representing a random graph. 
In particular, the value $x_{ij}^{(t)}$ is equal to one if an edge from node $i$ to node $j$ is present at time $t$, and it is equal to zero otherwise.
We obtain the matrices $\textbf{X}^{(1)}, \dots, \textbf{X}^{(T)}$ by thresholding at $0.1$ the partial correlations introduced in the previous section, that is, we observe an edge between two financial institutions if and only if the corresponding partial correlation is greater than $0.1$ in absolute value.

We choose the threshold value $0.1$ since this makes the graph densities very close to $0.5$ for all time frames. 
The least connected bank has $11$ neighbours, whereas the most connected bank has $23$ neighbours\footnote{These are the extreme values observed across all time frames.}, suggesting a fairly regular structure.
We aim at an ideal density of $0.5$ because this would give us a balanced scenario where we have sufficient information on the neighbourhood of each bank. 
More importantly, the literature on LPMs suggests that high density graphs are more suitable for latent space modelling, since edges generally provide more information than non-edges when we are interested in inferring latent positions \cite{raftery2012fast,rastelli2018computationally}.

In order to check the sensitivity to the value $0.1$, we also ran all the LPM simulations with threshold value $0.225$ (this leads to graph densities close to 0.2) and $0.085$ (this leads to graph densities close to 0.65): we did not notice any substantial qualitative change in the results\footnote{Results available upon request.}.


\subsubsection{Model}\label{sec:model}
LPMs can be considered as generative models for the presence vs absence of edges in a random graph.
By contrast, in this paper we are interested in modelling the log default probabilities associated to the nodes of the graph.
Our modeling assumption postulates that, at each time, each institution is characterised by a vector of latent coordinates $\textbf{z}_{i}^{(t)} \in \mathbb{R}^2$.
These values are model parameters which must be estimated from the observed data. 
We construct our approach following two core ideas:
\begin{enumerate}
 \item {\color{black}I}nstitutions located close to each other will tend to exhibit similar default probabilities, that is the risk on one node will have a certain influence on the risks of the other nearby nodes.
 \item {\color{black}T}he intensity of this contagiousness is determined by the Euclidean distance between the institutions, and by whether the institutions interact with each other or not.
\end{enumerate} 

We introduce a new dynamic LPM to model the log default probabilities in the time periods considered.
We assume that, conditionally on the latent positions, the likelihood function has the following form:
\begin{equation}\label{eq:likelihood_1}
 f\left( \textbf{y} \middle\vert \textbf{Z}, \textbf{X} \right) 
 \propto \exp\left\{ - \frac{1}{2}\sum_{t=1}^{T} \sum_{i\neq j} x_{ij}^{(t)} \eta_{ij}^{(t)} \left( y_i^{(t)} - y_j^{(t)} \right)^2\right\}.
\end{equation}
In the above equation, the summation is over all the pairs of $i$, $j = 1,\dots,N$ such that $i \neq j$; 
$\eta_{ij}^{(t)}$ is a similarity measure and it corresponds to one over the Euclidean distance between the nodes.

The following proposition motivates the likelihood definition in \eqref{eq:likelihood_1}.
\begin{proposition}\label{prop:gaussian_1}
The full conditional $\pi\left( y_i^{(t)} \middle\vert \textbf{y}_{-(i,t)}, \textbf{Z}, \textbf{X} \right)$ is a Gaussian distribution with mean $\mu_i^{(t)}$ and variance $\nu_i^{(t)}$ as follows:
\begin{equation}\label{eq:prop_1}
\mu_i^{(t)} = \frac{ \sum_{j=1}^{N} x_{ij}^{(t)} \eta_{ij}^{(t)} y_j^{(t)} }{ \sum_{j=1}^{N} x_{ij}^{(t)} \eta_{ij}^{(t)} },
\quad \quad \quad
\nu_i^{(t)} = \frac{1}{ \sum_{j=1}^{N} x_{ij}^{(t)} \eta_{ij}^{(t)} }.
\end{equation}
\end{proposition}
The symbol $\textbf{y}_{-(i,t)}$ indicates the collection of all observed data with the exception of the $t$-th value of bank $i$.
The proof of the proposition is shown in Appendix \ref{app:gaussian_1}.
This result underlines a straightforward interpretation for the model: 
conditionally on all parameters being fixed, the expected log default probability of an institution is equal to the weighted average of the log default probabilities of its graph neighbours, with weights corresponding to the similarity measure given by $\eta_{ij}$. 
This interpretation is in agreement with the first idea described in Section \ref{sec:model}.

In practice, we add a tiny quantity $\varepsilon = 0.001$ to the denominators of \ref{eq:prop_1} to avoid degeneracy whenever a node has no neighbours. 
With this modification, the full conditional for a node with no neighbours is a Gaussian centred in zero and with a very large variance, which is reasonable since it gives substantial flexibility on the realisation of the corresponding value.
If $\varepsilon$ is small enough, the results are not affected.

\subsubsection{Hierarchical structure}\label{sec:structure}
We create a hierarchical structure and specify prior distributions on the latent positions.
This is in close agreement with the literature on LPMs, where Bayesian settings are most commonly used.
We consider a standard Gaussian prior on the latent positions, and on the innovations of the latent positions.
This means that $\textbf{z}_{i}^{(1)}$ follows a bivariate Gaussian centered in zero with the identity as covariance matrix; 
whereas, for $t>1$, $\textbf{z}_{i}^{(t)}$ follows a bivariate Gaussian centred in $\textbf{z}_{i}^{(t-1)}$ with the identity as covariance matrix.

The prior distributions that we use in our setup may be regarded as informative.
As we will discuss in the following sections, our framework relies on an optimisation approach:
in this perspective, the prior distributions act as penalisations for the log-likelihood.
As a consequence, we do not use these directly to model any prior information that we have on the model parameters, but rather we use them to penalise degenerate scenarios and to emphasise solutions that are most relevant in our context.
In practice, we promote small innovations on the latent positions to guarantee that the latent network snapshots remain comparable across times and to facilitate the interpretability of our results.

\subsubsection{Inference and model interpretation}\label{sec:inference}
The likelihood function which is defined in Equation \ref{eq:likelihood_1} specifies the density kernel only up to a proportionality constant. 
The associated normalising constant does not have an analytical form, and is generally difficult to approximate numerically since it involves the calculation of a $NT$-dimensional integral.
Hence, in our approach we deal with a so-called \textit{intractable likelihood} \cite{moller2006efficient}, which creates a connection with a vast literature that deals with similar problems \cite{friel2012estimating}.

In a Bayesian setting, an intractable likelihood leads to a so-called doubly intractable problem \cite{murray2012mcmc}, whereby a standard implementation of a Markov chain Monte Carlo sampler may not be used efficiently.
To circumvent this limitation, we employ a pseudo-likelihood approximation:
\begin{equation}
 f\left( \textbf{y} \middle\vert \textbf{Z}, \textbf{X} \right) 
 \approx g\left( \textbf{y} \middle\vert \textbf{Z}, \textbf{X} \right) 
 = \prod_{t=1}^{T} \prod_{i=1}^{N} \pi\left( y_i^{(t)} \middle\vert \textbf{y}_{-(i,t)}, \textbf{Z}, \textbf{X} \right).
\end{equation}
We assume that the likelihood factorises into the product of the full conditionals of the data points.
Thanks to Proposition \ref{prop:gaussian_1}, we know that each of these full conditionals is proportional to a Gaussian, and hence we are able to compute the pseudo-likelihood exactly and efficiently.

As concerns parameter estimation, we propose to approximate the maximum-a-posteriori estimator using a simulated annealing scheme.
Simulated annealing can be seen as a stochastic optimisation algorithm that converges to a maximum of the objective function \cite{andrieu2003introduction}.
In our setup, the objective function is the pseudo-posterior which can be written as follows:
\begin{equation}\label{eq:pseudoposterior}
\tilde{\pi}\left( \textbf{Z} \middle \vert \textbf{X}, \textbf{y} \right)
\propto g\left( \textbf{y} \middle\vert \textbf{Z}, \textbf{X} \right) \pi\left( \textbf{Z} \right),
\end{equation}
where $\pi\left( \cdot \right)$ denotes a generic prior distribution.

The algorithm tries to update each of the parameters of the model in turn, by obtaining approximate samples from a tempered distribution.
When an update of a latent position is attempted, the new value $\textbf{z}_{new}$ is sampled from a bivariate Gaussian centered in $\textbf{z}_{i}^{(t)}$ with identity covariance matrix.
Let $\tilde{\pi}$ be the current value of the objective function, and $\tilde{\pi}_{new}$ be the new value of the objective after the change.
Then, the proposed update is retained with probability $\min\{1,\ ( \tilde{\pi}_{new} / \tilde{\pi} )^{1/\tau_k}\}$,
where $\tau_k$ is the tempering value, which decreases to zero during the procedure.

The optimisation approach allows us to speed up the inferential procedure, and to bypass the likelihood unidentifiability issues that are known to arise with LPMs \cite{shortreed2006positional}.
However, differently from other computational Bayesian approaches, the optimisation setting does not permit an assessment of the uncertainty around the point estimates that we obtain.

\begin{figure}[htb]
 \centering
 \includegraphics[width = 0.5\textwidth]{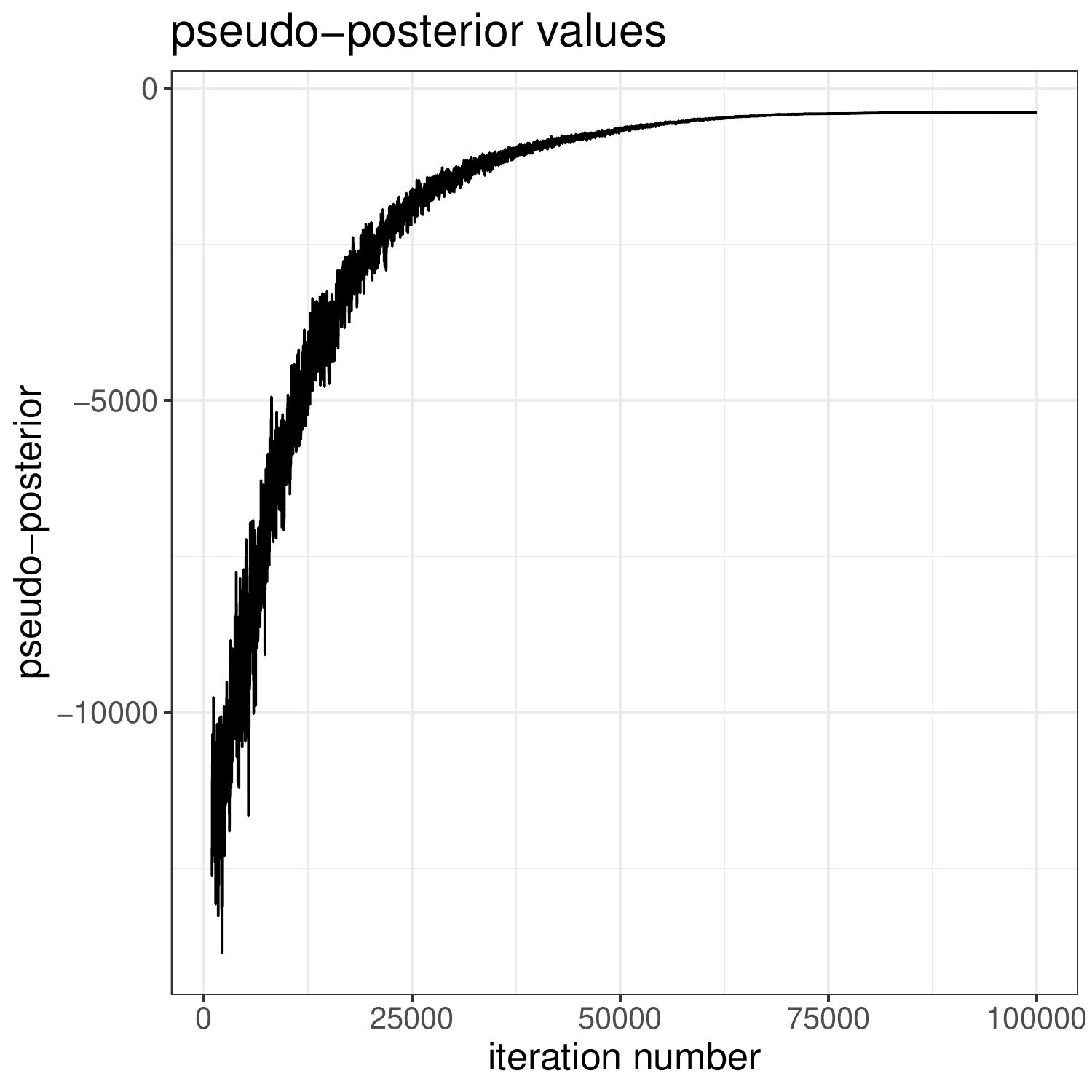}
 \caption{Pseudo-log-posterior values during the simulated annealing optimisation. The first $1{,}000$ iterations are not show for better clarity, since they correspond to a completely random update for the process.}
 \label{fig:sann_convergence}
\end{figure}

We run our algorithm on the dataset for $K = 100{,}000$ iterations, and temperature values defined by $100e^{-9.21 k/K}$, where $k$ is the iteration index.
The algorithm ran in $1.5$ hours on a $4$-cores machine. Figure \ref{fig:sann_convergence} illustrates the evolution of the calculated objective function during the optimisation.

A large variety of cooling schedules have been employed for simulated annealing algorithms \cite{winkler2012image}. The efficiency of these are highly context dependent, whereby users try out different values of the cooling function until good results are obtained. Generally, the cooling functions exhibit a decay which resembles that of the function that we adopt.Our cooling schedule hits a temperature value of $1$ (this value corresponds to the canonical Metropolis-within-Gibbs sampler) exactly half-way through.We emphasise that the total number of iterations $K$ is set to a very conservative value, since in our experiments we obtained the same qualitative results also for smaller $K$ values.

\section{Empirical results}\label{sec:results}The dataset that we study consists of the probabilities of default for 31 banks across 12 countries namely Austria, Belgium, Denmark, France, Germany, Italy, Netherlands, Norway, Spain, Sweden, Switzerland and the United Kingdom based on Thomson Reuters' structural model. Thomson Reuters evaluates the equity market's view of credit risk via a proprietary structural default prediction framework. Thomson Reuters produces daily updated estimates of the PD or bankruptcy within one year for more than 35,000 companies in the world, where the PDs are ranked to create 1-100 percentile scores, \cite{pourkhanali2016measuring}. We split it into four time periods which is shown in Table \ref{UniformT}.

\begin{table}[h]
\centering
\scalebox{1.1}{
\begin{tabular}{ccc}\hline
{Period} & {Start Date} & {End Date}\\\hline
Pre-crises & 3/Jan/2005 & 2/Jan/2008 \\
Financial Crisis & 3/Jan/2008 & 2/Jun/2010 \\
Sovereign Crisis & 3/Jun/2010 & 2/Jan/2013 \\
Post-crises & 3/Jan/2013 & 17/Nov/2016 \\
\hline 
\end{tabular}}
\caption{Explanation  of {\color{black} four different} periods in data set.}
\label{UniformT}
\end{table}

\subsection{Descriptive Statistics}
A summary of the values of default probabilities for each bank is provided in Table \ref{PDbyBank} in Appendix. Banks were also categorised according to their country of origin. The default probabilities for each country were then obtained by averaging the default probability values across the appropriate banks. A summary of these values are provided in Table \ref{PDbyCountry} in Appendix. 
Figure \ref{BoxPlot} shows the boxplots for the logarithm of probabilities of default of 31 banks from 2005-2016. {\color{black}  Due to having long tails, we used $\log (PD)$ to have more visible Boxplots.} From these plots, it can be inferred that certain banks are more volatile and that their default probabilities change more drastically. These include Banca Monte dei Paschi di Siena (BMPS), Commerzbank AG and ING Group. 

 \begin{figure}[h!]
\centering
   \begin{subfigure}{0.49\textwidth}
                \includegraphics[width=1\textwidth]{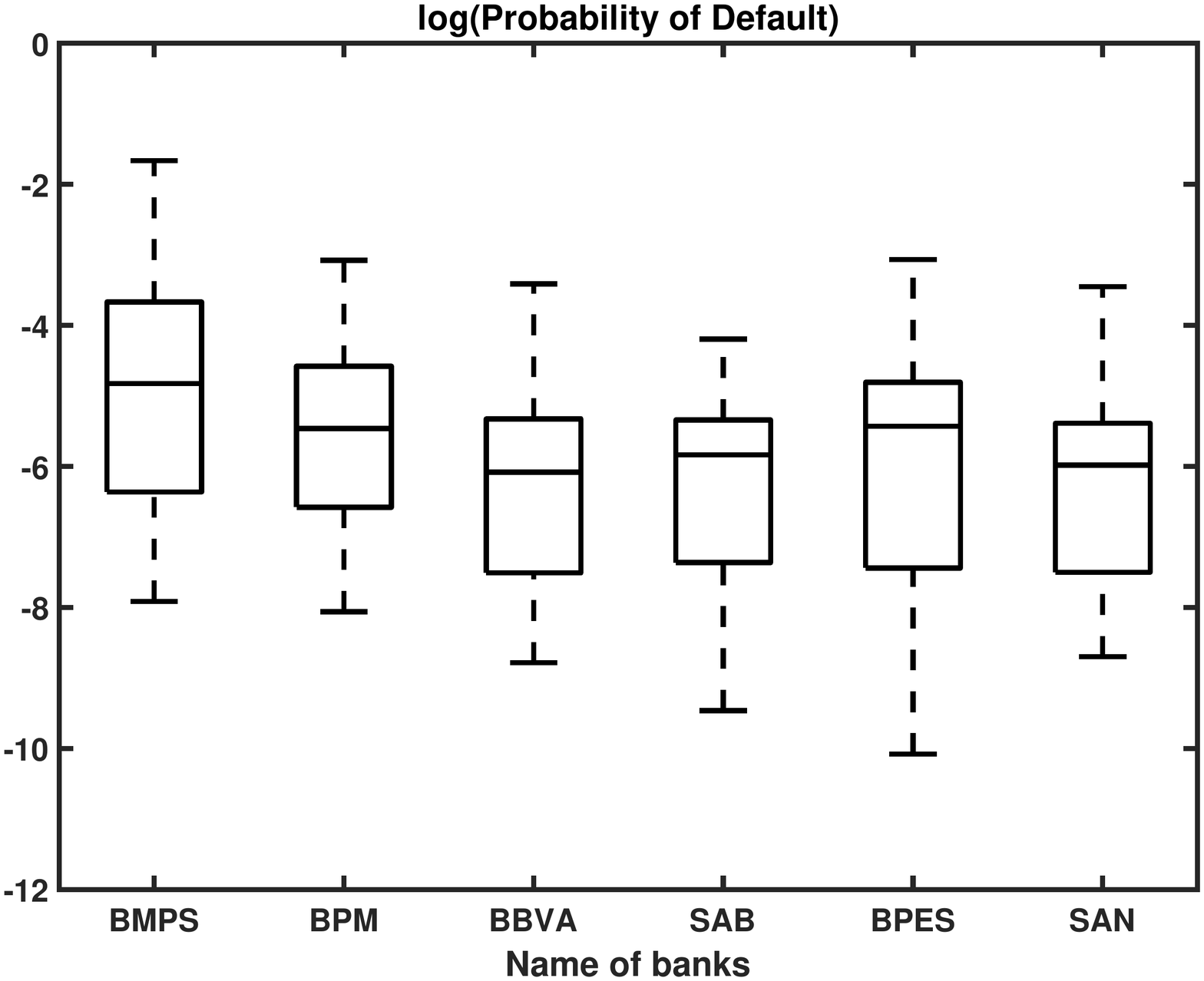}
        \end{subfigure}
        ~ 
        \begin{subfigure}{0.49\textwidth}
                \includegraphics[width=1\textwidth]{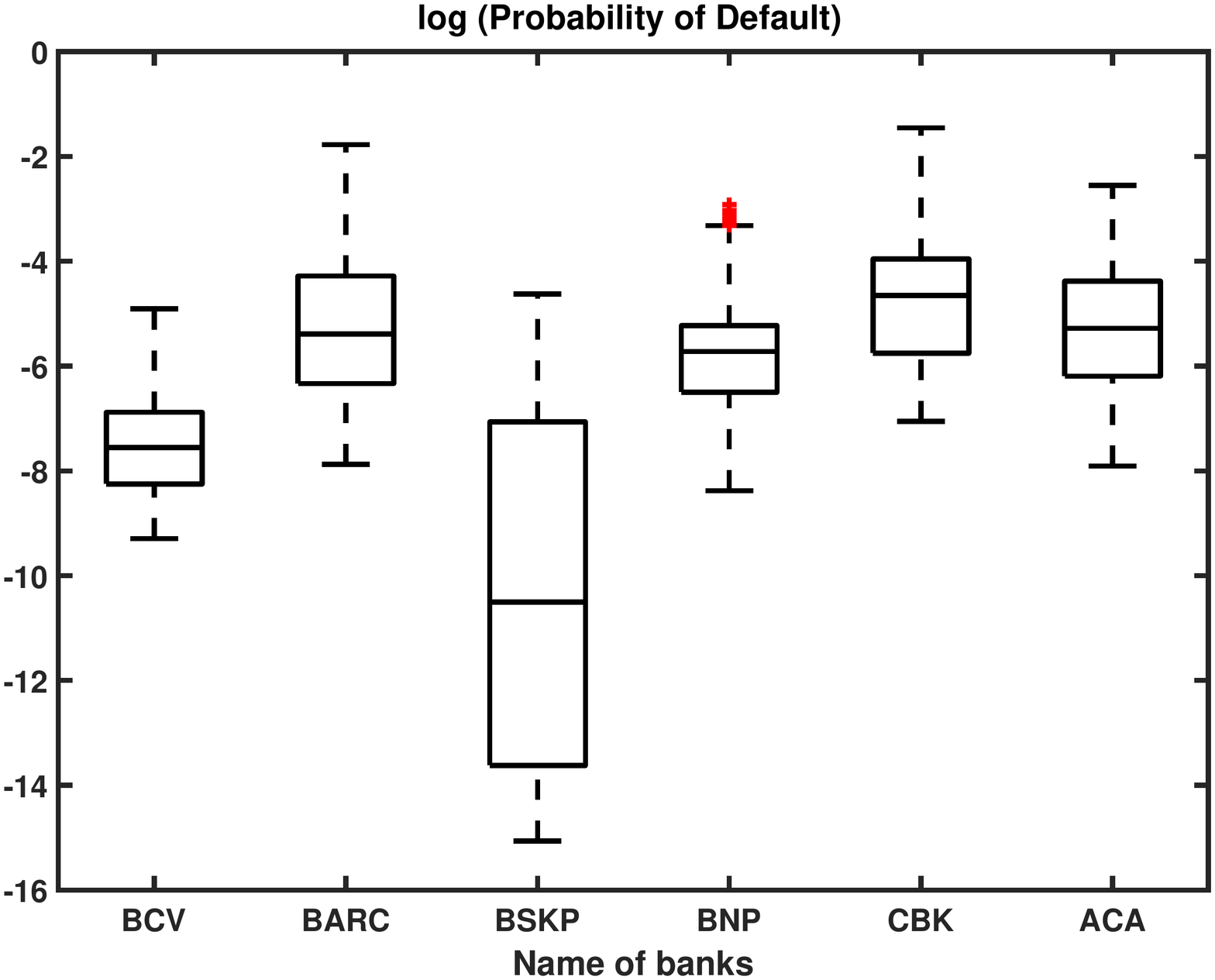}
        \end{subfigure}
\\
        \begin{subfigure}{0.49\textwidth}
                \includegraphics[width=1\textwidth]{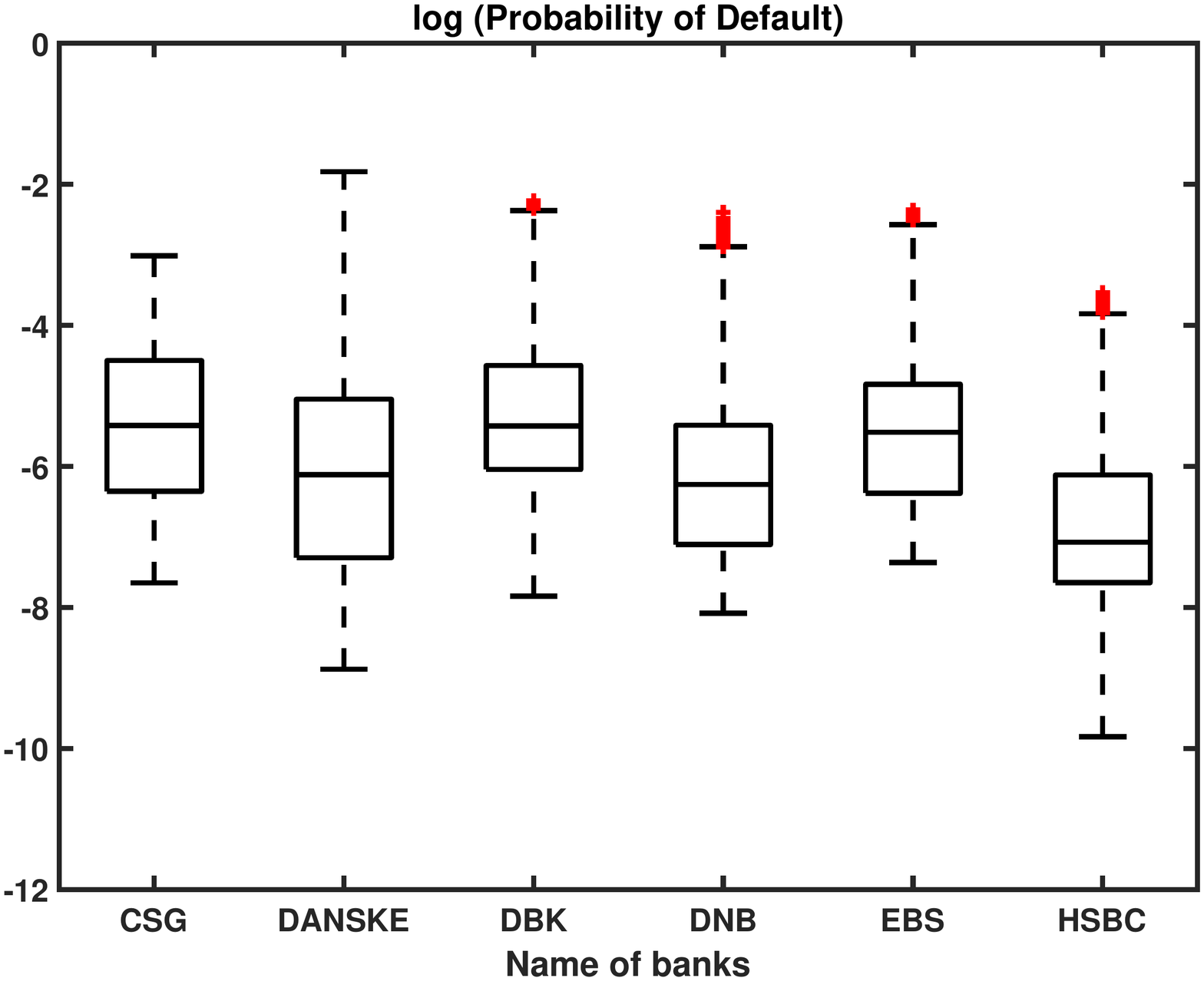}
        \end{subfigure}
        ~ 
        \begin{subfigure}{0.49\textwidth}
                \includegraphics[width=1\textwidth]{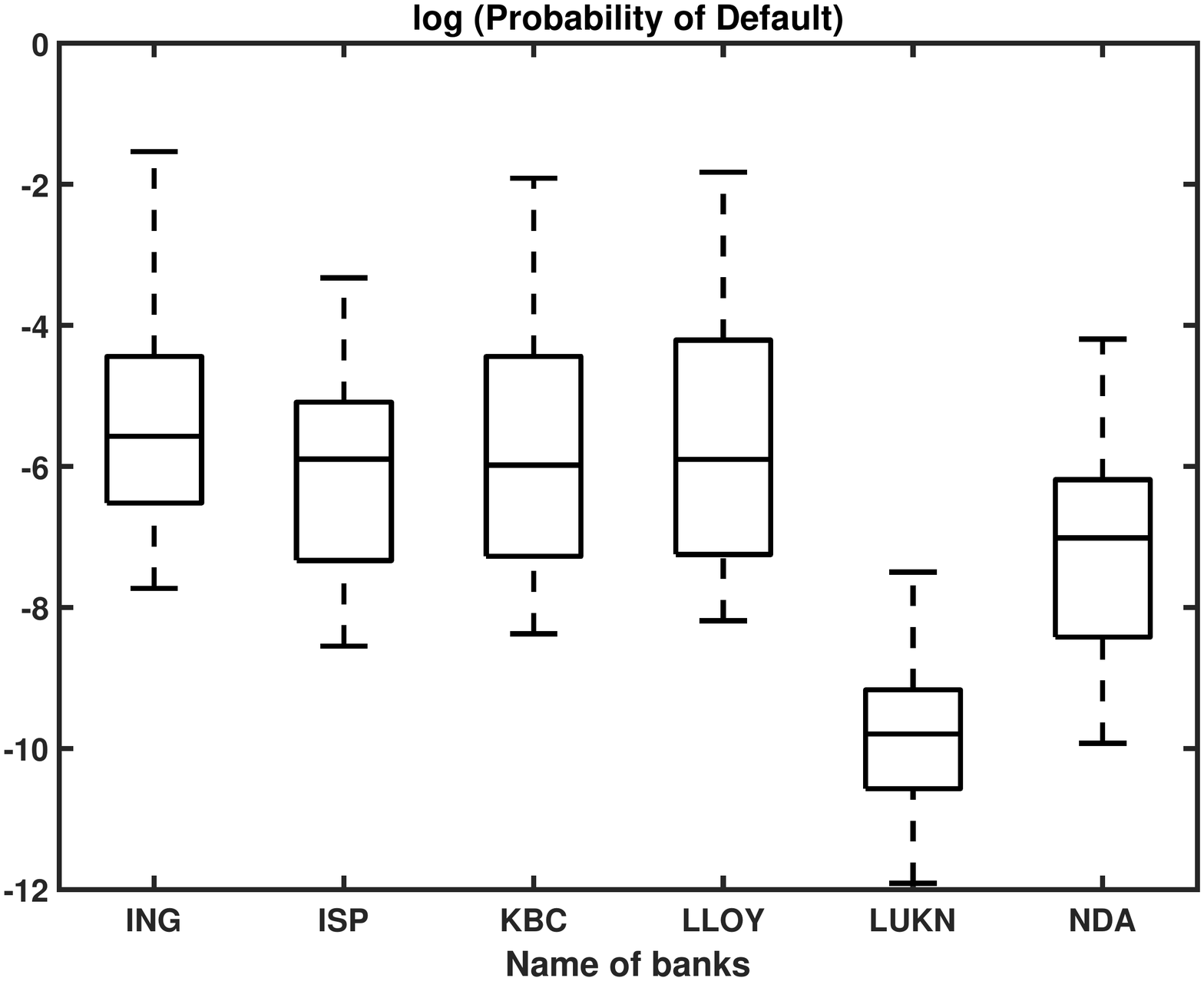}
        \end{subfigure}
\\
        \begin{subfigure}{0.49\textwidth}
                \includegraphics[width=1\textwidth, height=6 cm]{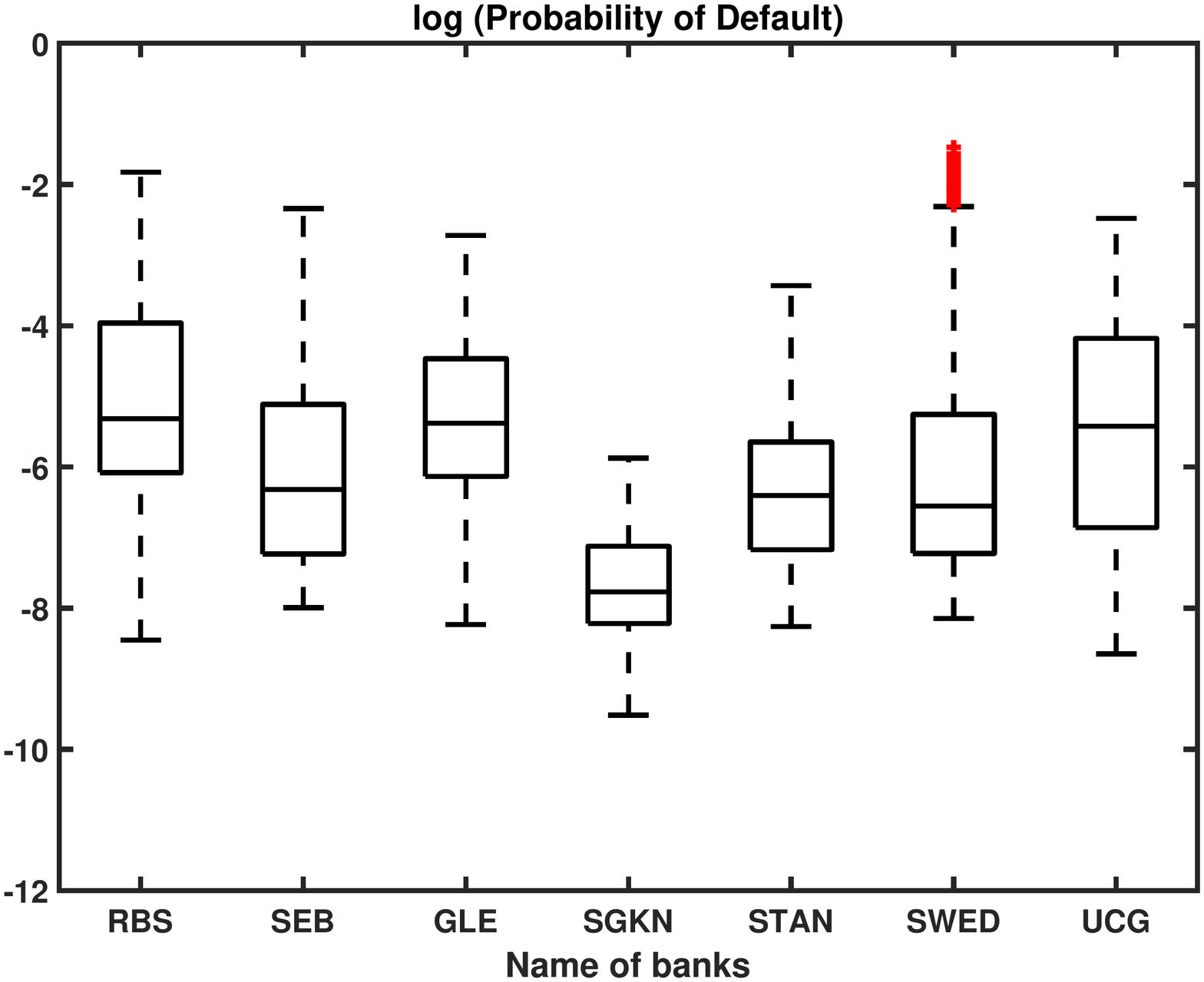}
        \end{subfigure}
        \caption{Boxplot of log (PD) for 31 banks.}
\label{BoxPlot}
\end{figure}
\subsection{Partial Correlation}
To determine the connectedness between the banks during each period, the partial correlation values between bank $i$ and bank $j$ were calculated and tested for significance using the \texttt{R} statistical software.\footnote{Results are available upon request.} 

\begin{figure}[htb]
 \centering
 \includegraphics[width=.9\textwidth, height=10cm]{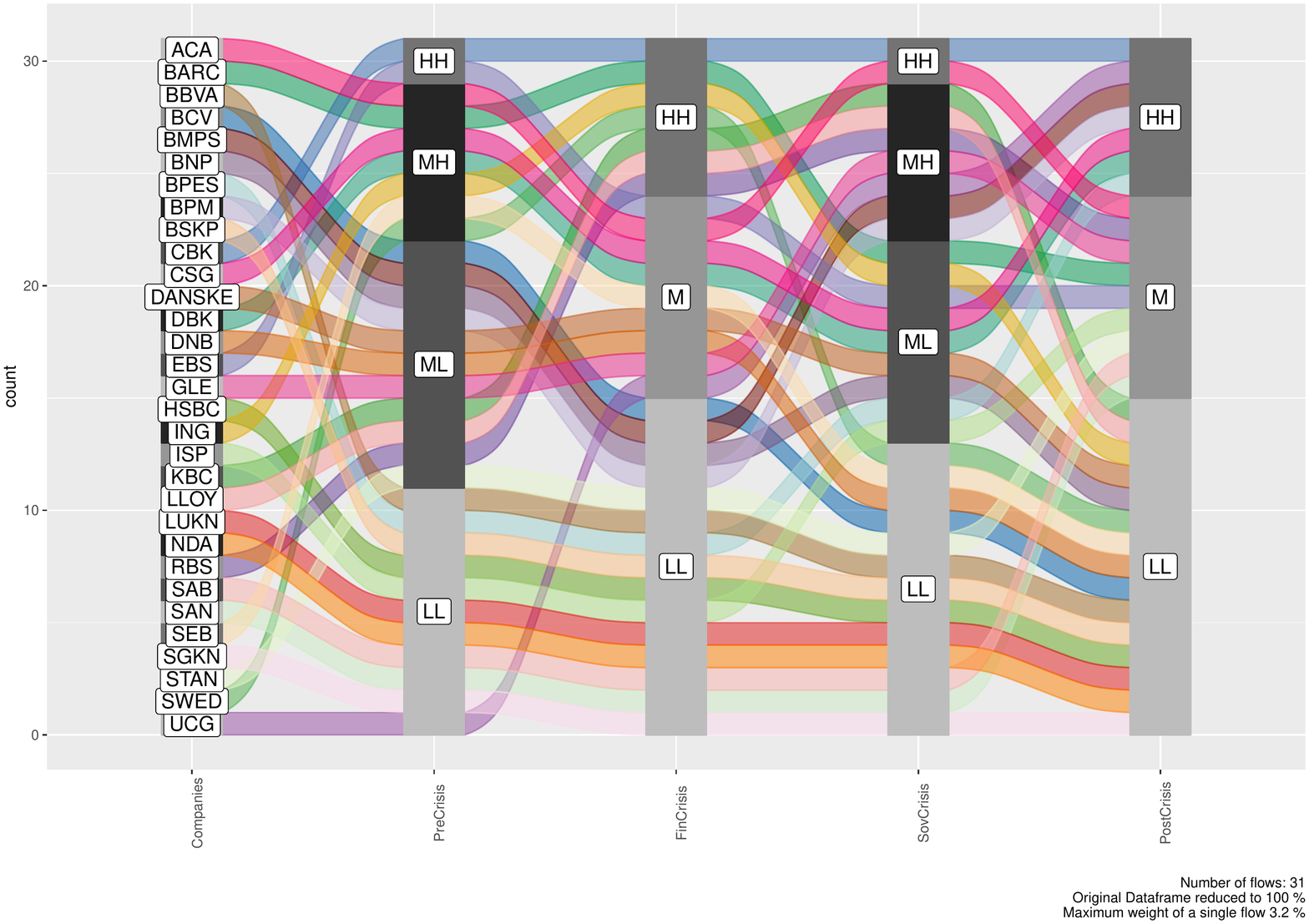}
 \caption{{\color{black}Alluvial diagram illustrating the changes of PD for banks over the four different periods. Bins are labeled as ‘LL’ (low-low), ‘ML’ (medium-low), ‘M’ (medium), ‘MH’ (medium-high) and ‘HH’ (high-high) by default.}}
 \label{fig:Allu}
\end{figure}

Our naive assumption is that during the financial and sovereign crises, the connections between banks would be closer together. Therefore, we would expect more significant partial correlations during these periods. The table of counts below demonstrates that our assumption is true during the financial crisis, but not during the sovereign crisis. However, this result alone is not sufficiently strong as it does not take into account the actual value of the partial correlations. \\
\begin{table}[H]
\centering
\begin{tabular}{cccc}
\hline
Precrisis & Financial Crisis & Sovereign Crisis & Postcrisis \\
\hline\\
273 & 285 & 271 & 241 \\
\hline
\end{tabular}
\caption{Number of Significant Partial Correlations.}
\label{tab:sigparcor_count}
\end{table}
{\color{black} In order to have a better visual map of ranking changes of banks, Figure. \ref{fig:Allu} is plotted here. It shows how the PD of banks and financial institutes (eventually ranking of banks) change over subsequence periods.}
\subsection{ CoRisk  Values}

\begin{figure}[htb]
   \begin{tabular}{@{}cc@{}}
      \includegraphics[width=.5\textwidth, height=7cm]{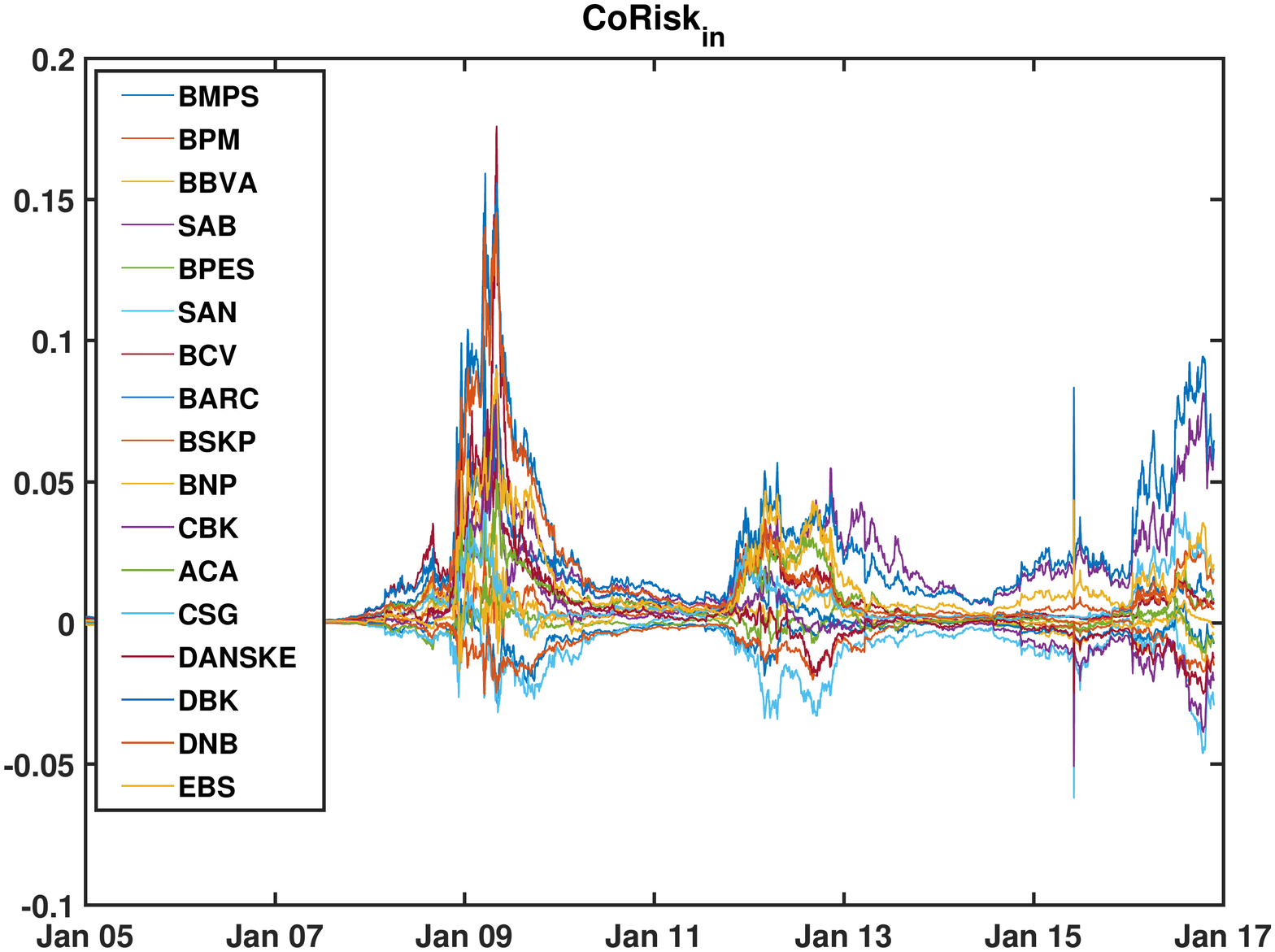} &
 \includegraphics[width=.5\textwidth, height=7cm]{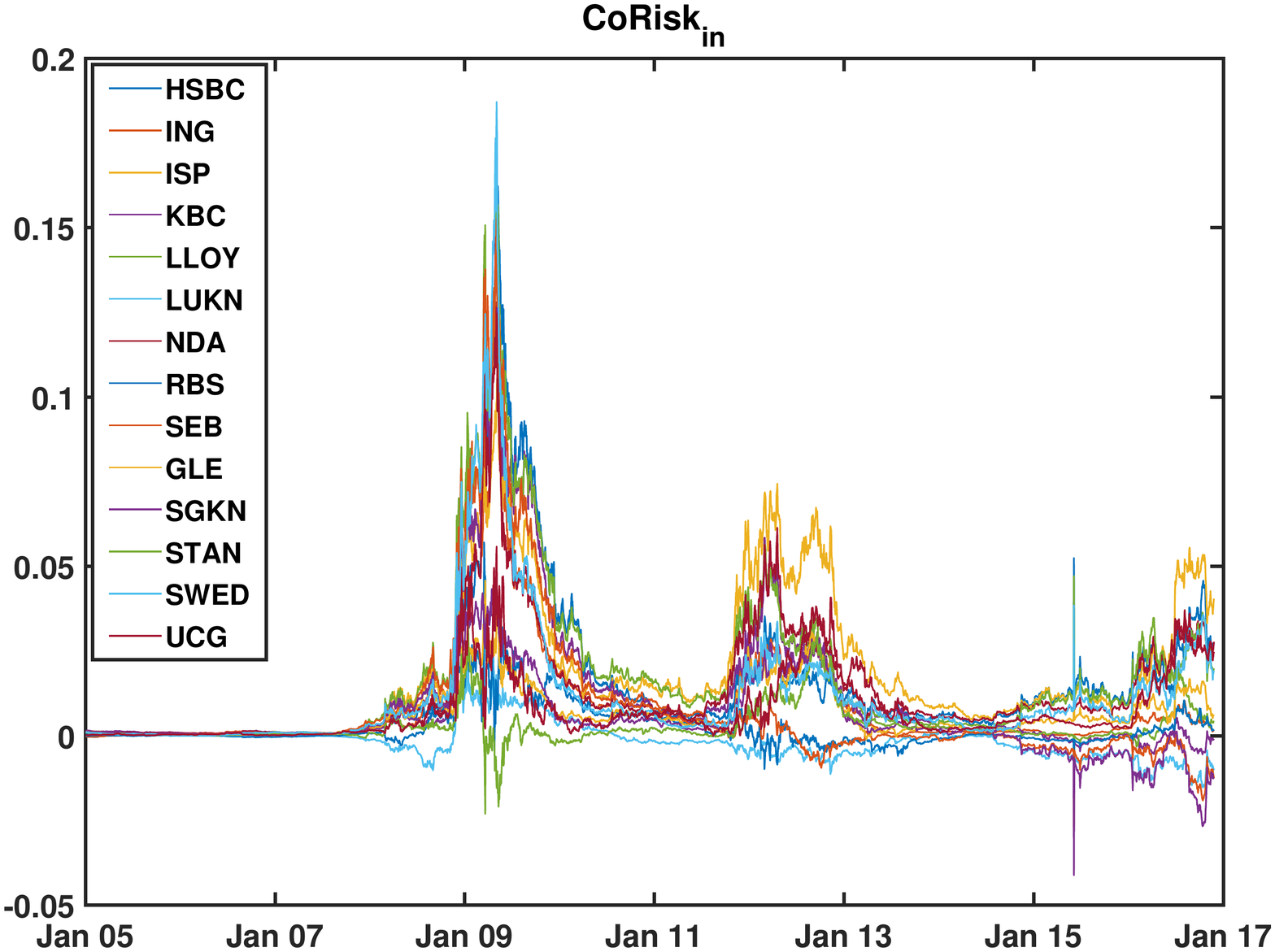} \\
 \includegraphics[width=.5\textwidth, height=7cm]{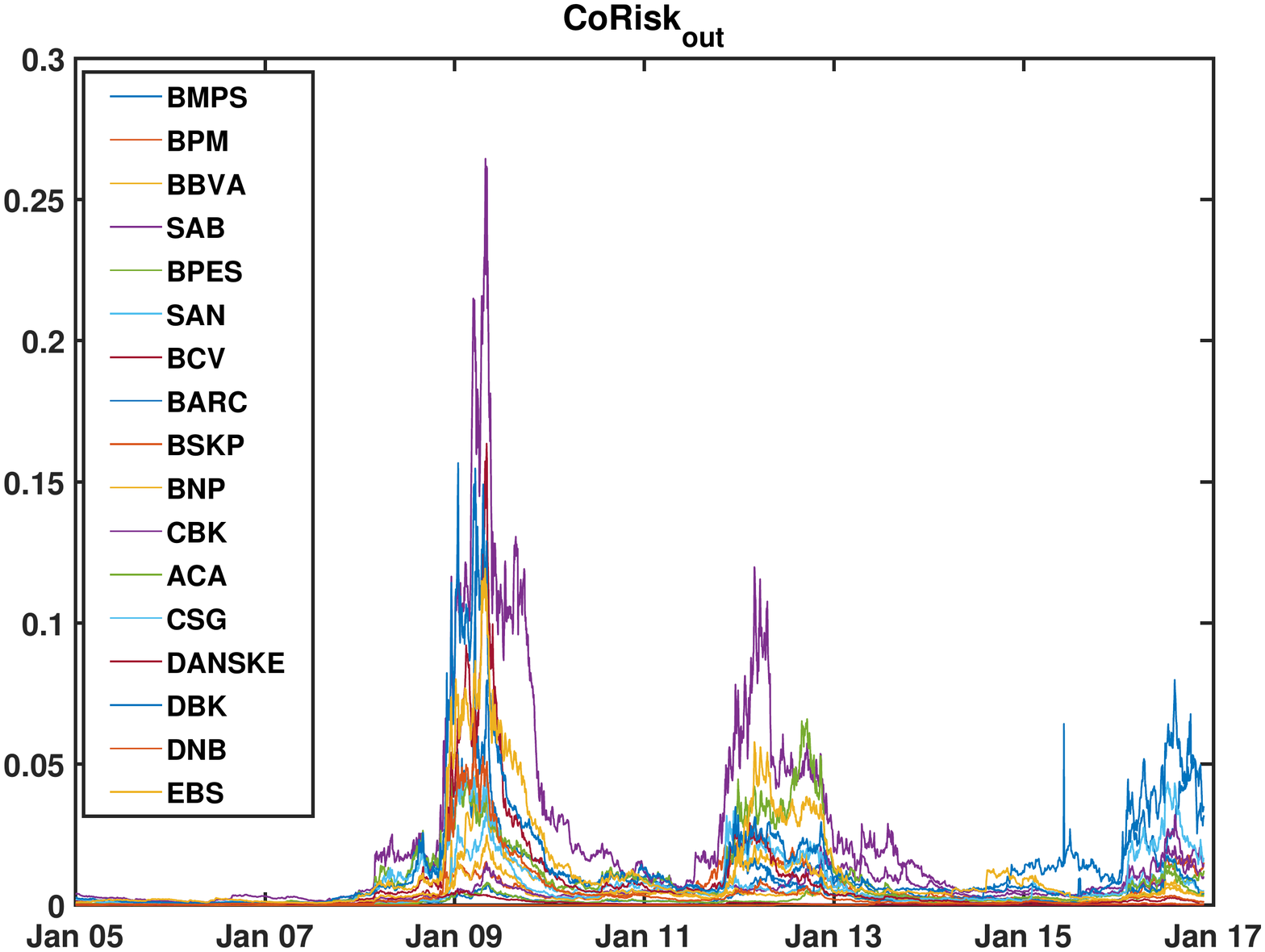} &
 \includegraphics[width=.5\textwidth, height=7cm]{CoRiskOut1.pdf} \\
  \end{tabular}
\caption{Time series of  CoRisk$_{In}$ and  CoRisk$_{Out}$ are shown in these plots for whole period of study.}
\label{CoRiskPlots}
 \end{figure}

After the calculation of the partial networks, and using the 1-year default probability values for each bank, we are now able to calculate the  CoRisk$_{in}$ and  CoRisk$_{out}$ at each point in time. The change in  CoRisk  values are shown in Figure \ref{CoRiskPlots}. {\color{black}To have more clear plots, we splitted data into two different groups and $CoRisk_{in}$ and $CoRisk_{out}$ are plotted for each group.} These figures exhibit spikes in  the  CoRisk  values during the financial and sovereign crises, confirming that more risk was transmitted during these periods. The  CoRisk  values after $2012$ remain higher than they were before the crises with smaller sudden spikes. This could be due to the lingering effects of the crises and the fact that some banks are yet to stabilise. 
The  CoRisk$_{ij}$ matrix was also calculated based on the significant partial correlation matrix obtained earlier for each period\footnote{Results are available upon request.}. 

\subsection{Financial contagion tests}

In order to see if there was a significant increase (or decrease) in the CoRisk$_{ij}$ value between different periods, paired t-tests (the $31 \times 31$  CoRisk$_{ij}$ matrices were transformed into a vector of 961 elements) were conducted between two periods. We found that there is a significant increase in the  CoRisk$_{ij}$ values during the financial crisis and also post-crises as compared to pre-crises. The results of our paired t-tests are summarised in Table \ref{tab:pairedttest}.\\

\begin{table}[H]
\centering
\scalebox{1.1}{
\begin{tabular}{cc}
\hline
 CoRisk$_{ij}$ &{p-value} \\\hline
financial crisis $>$ pre-crises & $0.0007$ \\ 
pre-crises $>$ post-crises & $0.02321$ \\
sovereign crisis $>$ financial crisis & $0.0010$ \\
financial crisis $>$ post-crises & $0.0078$ \\
\hline
\end{tabular}
}
\caption{Paired t-test.}
\label{tab:pairedttest}
\end{table}
\subsection{Kendall's tau test}
We are interested in seeing if the dependency of default probabilities between banks and countries increased from the pre-crisis period to financial crisis period. We use Kendall's tau measures the dependency between variables.  In order to construct the test, sampling with a replacement method was employed. Each column was re-sampled (with replacement) 200 times for the calculation of Kendall's tau. A t-test is then performed to see if $\tau$ has increased. At a significant level of 5 percent, only a few pairs of countries were identified as having a significant increase in dependency. They are: Spain /Switzerland, Germany/Denmark, UK/Norway. Since our crisis period was further split into two groups as financial crisis and sovereign crisis, we then combined them to see if the pre-crisis period would have a lower $\tau$ compared to the combined crisis periods. However, only Switzerland/Germany and Switzerland/Norway showed a significant increase in $\tau$. To confirm the robustness of this test, the re-sampling process was done a couple of times and each time the result varied. Hence we conclude that this would not be an effective way of determining the existence of contagion. In addition, among all 66 pairs of countries, the small proportion of significant pairs is another indicator that we need to explore further into the way of determining contagion effect. \\
The same procedure was repeated for banks. However, due to the uncertainty discussed above, we cannot draw any conclusions about the existence of contagion based on Kendall's tau test. Therefore, we apply another approach in the next section which is called the Minimum Spanning Tree.

\subsection{Minimum Spanning Tree (MST)}

\begin{figure}[H]
\centering
\includegraphics[scale=0.38]{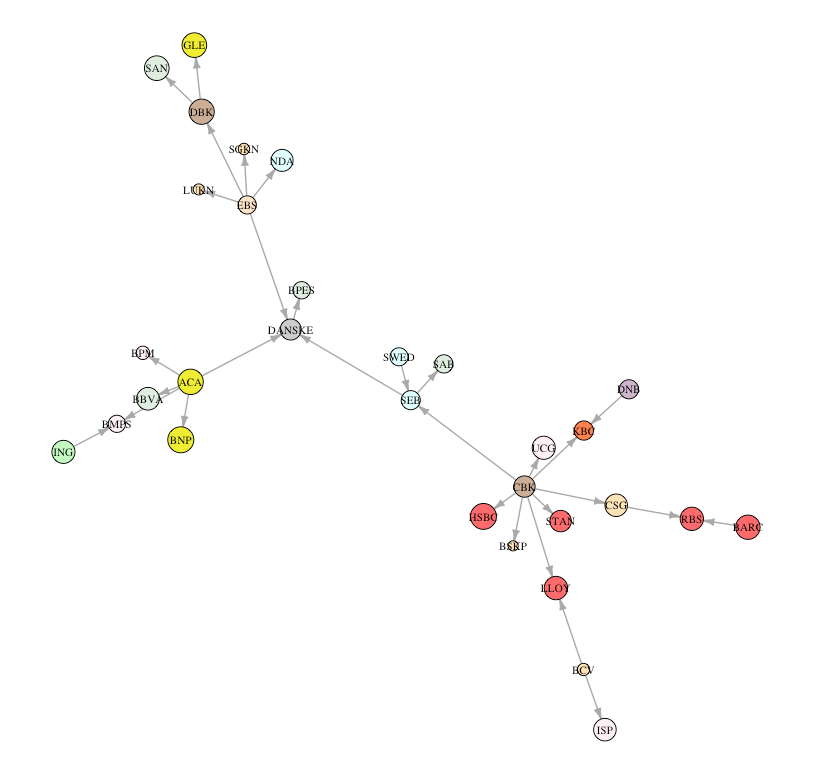}
\caption{Minimum Spanning Tree Pre Crisis by bank.}
\label{figMST1}
\end{figure}

\begin{figure}[H]
\centering
\includegraphics[scale=0.38]{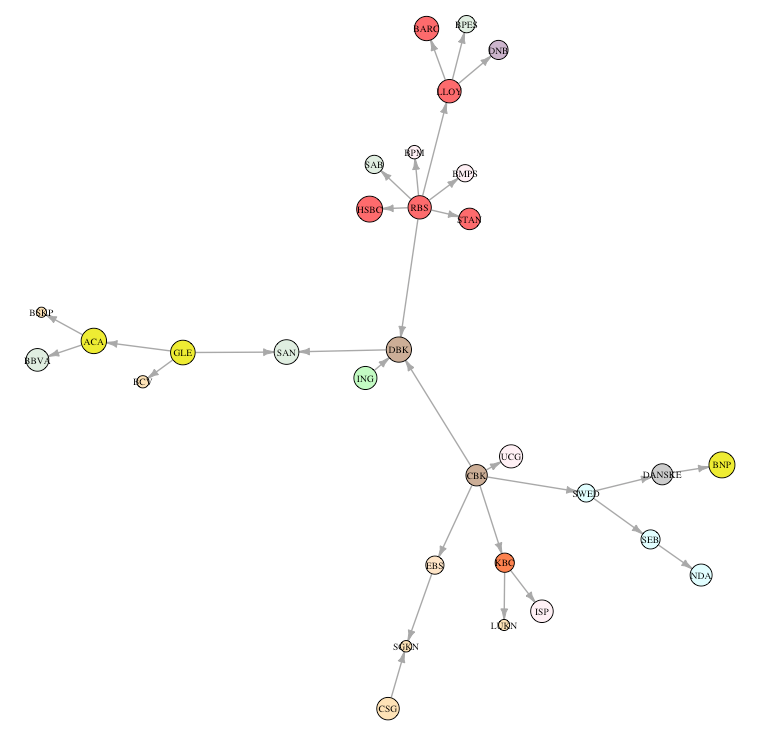}
\caption{Minimum Spanning Tree Financial Crisis by bank.}
\label{figMST2}
\end{figure}

\begin{figure}[H]
\centering
\includegraphics[scale=0.38]{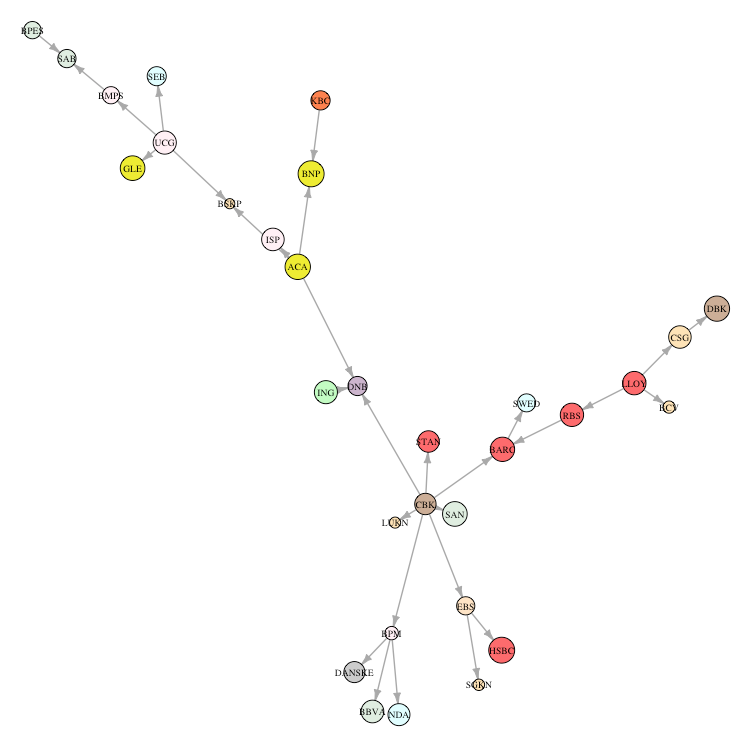}
\caption{Minimum Spanning Tree Sovereign Crisis by bank.}
\label{figMST3}
\end{figure}

\begin{figure}[H]
\centering
\includegraphics[scale=0.38]{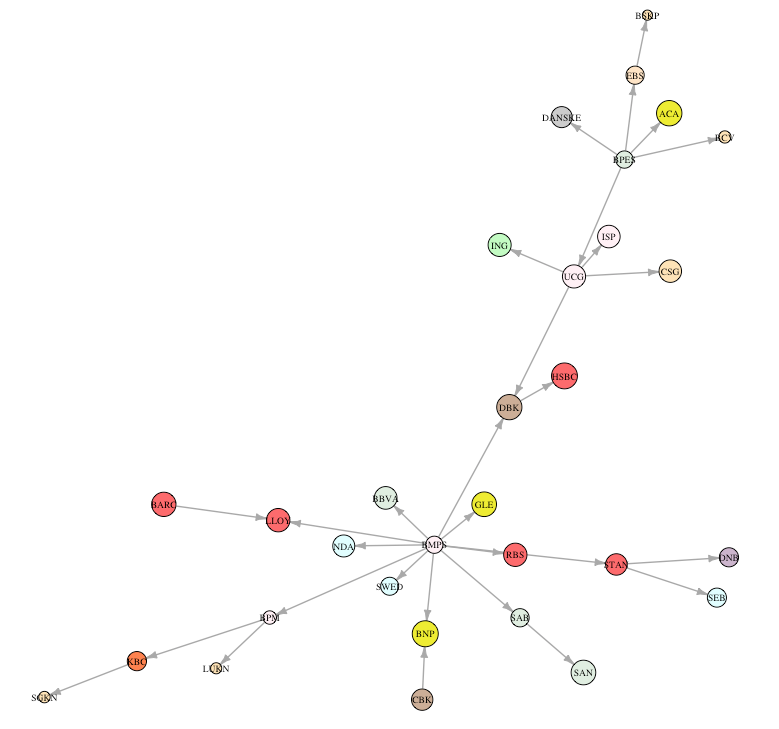}
\caption{Minimum Spanning Tree Post Crisis by bank.}
\label{figMST4}
\end{figure}

Network models are frequently adopted in the field of financial research due to their effectiveness at visualising large data sets. They have proven to be an effective resource in the prediction of market movements. One of the more popular methods for visualisation of financial networks and building the dependence network is the minimum spanning tree (MST) approach (Mantegna, 1999) which is designed to select or filter the information presented in the dependence (or correlation) matrix. The metric distance for creating the edges of a financial network, such as in  \cite{wang2018correlation}, uses the partial correlation values, $C_{ij}$, and is given by the following equation
\begin{eqnarray*}
d_{ij}=\sqrt{2(1-C_{ij})}.
\end{eqnarray*}
In our work, we use a different distance metric, replacing the partial correlation values with CoRisk  values. The formula for the distance between two edges will then be as follows,
\begin{eqnarray}
d_{ij}=\sqrt{2(1-CoRisk_{ij})}.
\end{eqnarray}
\begin{remark}
Unlike the partial correlation values, the  CoRisk  values are not symmetric, that is $d_{ij} \neq d_{ji}$. Since  CoRisk  takes values from $-\infty$ to $1$, $d_{ij}$ would always be positive. Furthermore, a larger  CoRisk$_{ij}$ value implies that bank $i$ has a greater impact on bank $j$, and this is represented by a shorter distance. 
\end{remark}
A directed graph consisting of 31 nodes (31 banks) is then formed using the distances obtained (a pair of directed edges between two banks is only present if the banks are significantly correlated as determined earlier). This directed graph consists of a large number of directed edges and is difficult to visualise. Therefore, we used the \texttt{R} software to obtain a MST with directed edges. The MST from each of the four periods are shown in Figures \ref{figMST1}-\ref{figMST4}.

{\color{black}
In this work, we consider MST approach based on Banks and countries, separately. Billio et al. \cite{billio2018networks} use a time varying weight matrix, but they specify that it can vary at a time scale lower than the entities in the system which is used.  We use a constant weight matrix in this work. When determining a weight matrix, it is important that this matrix reflects any connections between countries. One could argue a spatial weight matrix could fill this role of reflecting connectedness. The most simple form might be a binary weight matrix based on the existence of a neighbouring relationship. The results from this, however, are less than satisfactory, and a matrix that better shows the complex nature of economic relationships was desired. In this research, using European countries, the option of travel distance between capitals or stock market cities seems another option at first glance. Countries closer in distance may be more likely to carry out trade between themselves, linking their stock markets more than those far away. Flavin et al.\cite{flavin2002explaining} note that overlapping business hours tend to increase correlations between countries stock markets. Applying this logic in this data set, the inverse function of travel distance does not prove a reliable option because, with an increase in globalisation, it is fair to expect that travel distances will not be an efficient factor in stock market correlations. 
\begin{remark}
 In order to apply MST for Banks, they are colour-coded in accordance with their countries and made proportional to the value of their total assets in 2016. For example, Nordea is a Swedish bank coloured in yellow and had a total asset value of \euro 615 billion, while British bank Barclays is coloured in green and had a total asset value of \euro 1213 billion. The length of the edges between 2 banks is proportional to the distance metric used, and the direction shows which bank is transmitting the risk.
\end{remark}
}
\newpage

\begin{figure}[H]
 \centering
\includegraphics[scale=.376]{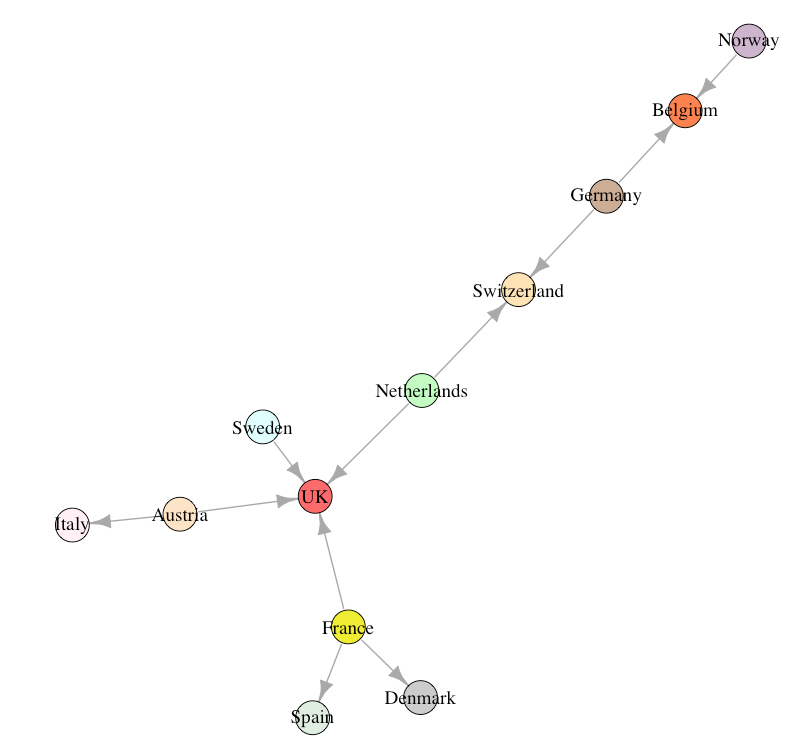}
\caption{Minimum Spanning Tree Pre Crisis by country.}
\label{figMST5}
\end{figure}

\begin{figure}[H]
\centering
\includegraphics[scale=.376]{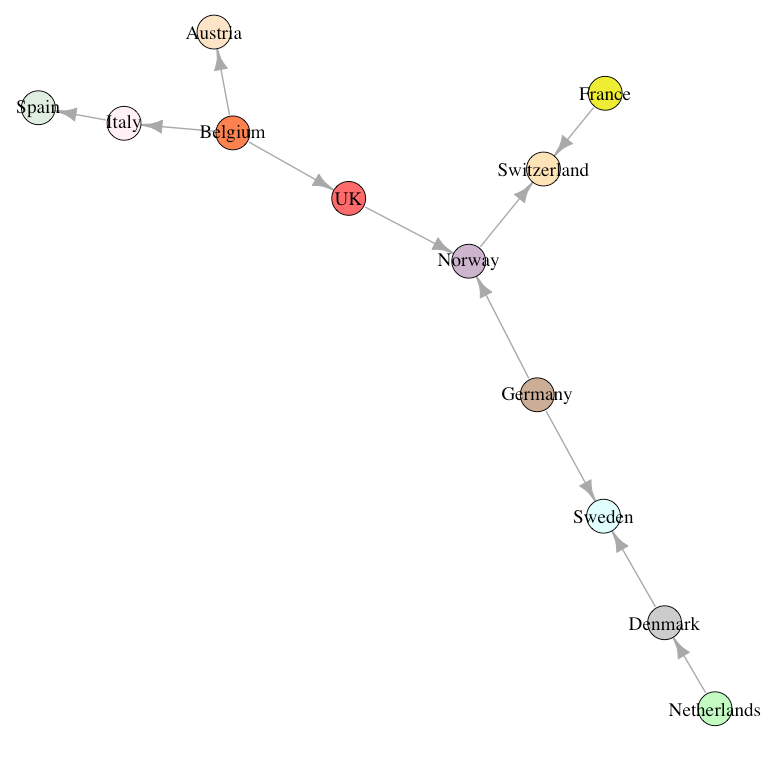}
\caption{Minimum Spanning Tree Financial Crisis by bank.}
\label{figMST6}
\end{figure}

\begin{figure}[H]
\centering
\includegraphics[scale=.376]{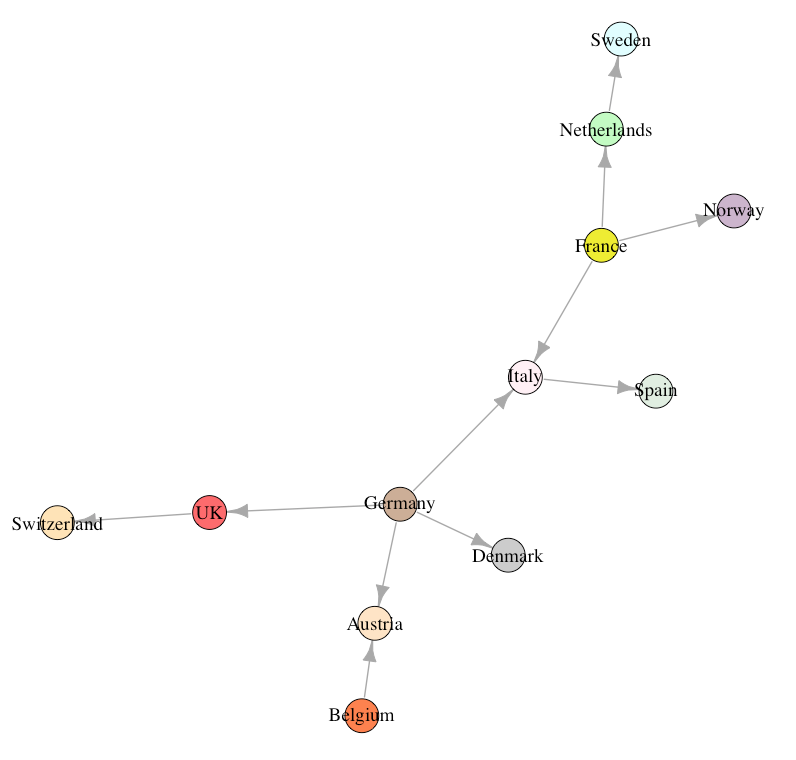}
\caption{Minimum Spanning Tree Sovereign Crisis by country.}
\label{figMST7}
\end{figure}

\begin{figure}[H]
\centering
\includegraphics[scale=.376]{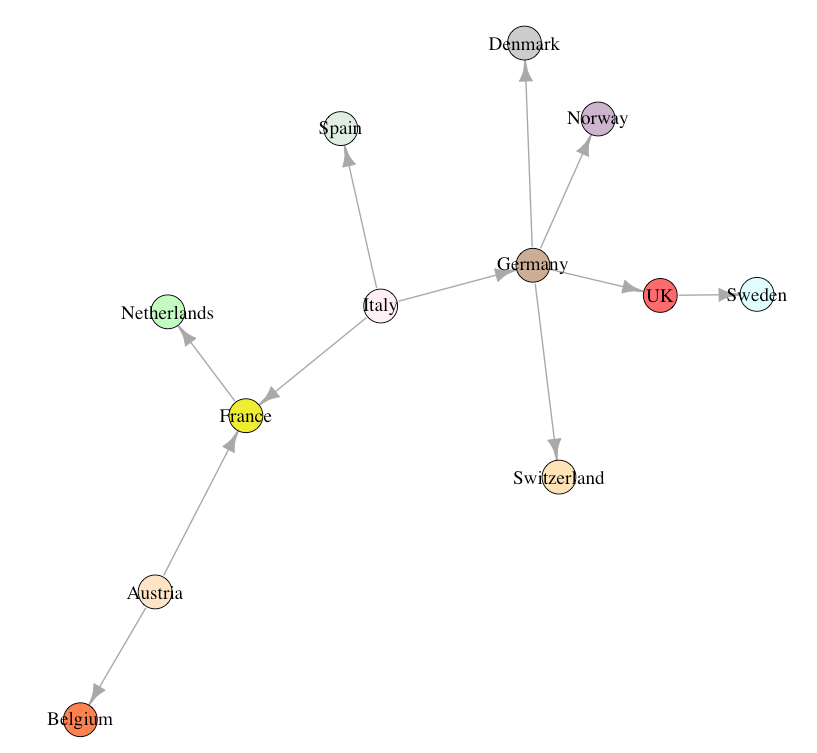}
\caption{Minimum Spanning Tree Post Crisis by country.}
\label{figMST8}
\end{figure}

Figure \ref{figMST1} shows the MST before the global financial crisis. The Swedish bank SEB is located in the middle of the MST and the other centrally located banks are Commerzbank AG, Danske Bank and Credit Agricole. During the financial crisis, the two German banks, Deutsche Bank and Commerbank AG are the centrally located banks in the network, as seen in Figure \ref{figMST2}. The banks also start to cluster according to their country, with the more notable ones being those from the United Kingdom and Sweden. This suggests that the banks within a country become more closely connected during the financial crisis. During the sovereign crisis, from Figure \ref{figMST3}, the clustering effect seems to be less apparent than during the financial crisis. Commerzbank AG remains a centrally located bank, along with Credit Agricole and DNB. Figure \ref{figMST4} shows that the most critically important bank post-crises is BMPS, where due to its high probability of default and financial instability following the 2 crises, it has the largest impact on other European banks. \\
The above MSTs represent the connections between banks. To determine the relationships between countries instead, MSTs were also plotted by considering the banks located in each country. These MSTs for each of the four periods are shown below. \\
The United Kingdom is the most centrally located country in the financial network before the financial crisis, as seen in Figure \ref{figMST5}. It remains centrally located during the financial crisis (Figure \ref{figMST6}) but the MST divides into two portions separated by the UK, one with Belgium and the other with Norway. During the sovereign crisis (Figure \ref{figMST7}), the major European countries, France, Germany and Italy became more centrally located. The structure of the MST remains fairly similar after the sovereign crisis (Figure \ref{figMST8}). In the next section, we analyse MSTs by using two approaches which are called as measures of centrality and fragility.

\subsubsection{Analysis of Minimum Spanning Tree using Fragility and Centrality measures}
Fragility, as proposed by \cite{das2016matrix}, is the propensity for risk to spread through a network.  This can lead to assess of the speed at which contagion can spread in the system. A network that has more links will transmit more risk as it spreads quicker throughout the network. Therefore, a network with a higher fragility score is more contagious. 

Fragility of a network can be described by the following equation, where $d$ represents the degree of a node in the network,
\begin{equation}
R = \frac{\mathbb{E}(d^2)}{\mathbb{E}(d)}.
\end{equation}

Fragility values were calculated for the bank networks and country networks across the four periods. The results are summarised in the Table below.

\begin{table}[ht]
\centering
\begin{tabular}{ccc}
\hline
Period & Bank Network & Country Network \\
 \hline
Pre-Crisis & $40.81$ & $16.75$ \\ 
 
Financial Crisis & $40.32$ & $16.58$ \\
 
Sovereign Crisis & $41.24$ & $19.26$ \\
 
Post-Crisis & $43.27$ & $18.86$ \\
\hline
\end{tabular}
\caption{Fragility Score.}
\label{tab:fragilityscore}
\end{table}
\begin{remark}
The fragility score gives us an overview of the transmission risk within the network, but fails to identify the sources of risk. Centrality measures are used to determine the main banks/countries that contribute to the source of the risk. 
\end{remark}
Below, we introduce some centrality measures used and the centrality scores of both banks and countries are then summarised.

The first centrality measure that we use in this work is Betweenness Centrality. Betweenness centrality for a node, $v$, in a network is defined as the number of shortest paths between two nodes that passes through node $v$. It is given by the following formula:

\begin{equation}
g(v) = \sum_{s \neq v \neq t} \frac{\sigma_{st}(v)}{\sigma_{st}}.
\end{equation}

A bank with a larger value for its betweenness centrality will be located in a more central location in the network and thus be a major transmitter of risk through the network. Betweenness values were calculated for each period for both the bank and country networks. 
\\\\
The second centrality measure which is called as closeness centrality, measures the inverse of the sum of distances from a node to all other nodes in a network. A higher closeness value suggests that a node is close to other nodes, (for more detail see \cite{grassi2010centrality}). Banks with a relatively high closeness value would mean that it would require less time for it to transmit risk to other banks in the network as these banks would be 'closer' to other banks in terms of the $CoRisk$ metric. The formula for closeness centrality is given by: 
\begin{equation}
closeness(v) = \frac{1}{\sum_{i \neq v}d_{vi} }.
\end{equation}
The third centrality measure is Laplacian centrality which calculates the drop in Laplacian energy (sum of squares of eigenvalues in a Laplacian matrix) when a vertex is removed, see \cite{Qi2013}. It incorporates the weights of the edges in the measure, which makes it a stronger measure than betweenness centrality. 
The centroid value is another centrality which measures the number of interactions a node has to determine how central the node is in the network. It can be used to determine which bank is the 'co-ordinator' within a network we use leaderRank. Banks with a relatively high centroid value will play a more central role in the financial network, \cite{scardoni2012centralities}.
In order to measure the influence of a node in a directed network, by using a standard random walk procedure, we use LeaderRank. Banks with a higher LeaderRank score are more influential in the financial network.

\begin{table}[b]
\caption{Centrality Scores for Banks: pre-crisis.}
\centering
\scalebox{0.99}{
\begin{tabular}{cccccc}
\hline
& Betweeness & Closeness & Laplacian & Centroid & LeaderRank \\ 
\hline
BMPS & 5 & 0.0177 & 3276 & -7 & 0.9913 \\ 
BPM & 3 & 0.0181 & 3534 & -9 & 1.0363 \\ 
BBVA & 6 & 0.0161 & 2348 & -15 & 0.8110 \\ 
SAB & 1 & 0.0181 & 3550 & -9 & 1.0363 \\ 
BPES & 2 & 0.0191 & {\textbf{4058}} & -9 & \textbf{1.1265} \\ 
SAN & 4 & 0.0172 & 3038 & -11 & 0.9462 \\ 
BCV & 12 & 0.0186 & 3776 & -9 & 1.0814 \\ 
BARC & 10 & 0.0177 & 3296 & -13 & 0.9913 \\ 
BSKP & 0 & 0.0181 & 3490 & -9 & 1.0363 \\ 
BNP & 11 & 0.0172 & 3026 & -7 & 0.9462 \\ 
CBK &\textbf{39} & 0.0181 & 3522 &\textbf{0} & 1.0363 \\ 
ACA &\textbf{36} & 0.0181 & 3510 & \textbf{-5} & 1.0363 \\ 
CSG & 15 & 0.0181 & 3498 & -11 & 1.0363 \\ 
DANSKE & 9 & 0.0168 & 2772 & -9 & 0.9012 \\ 
DBK & 15 & 0.0172 & 3026 & -7 & 0.9462 \\ 
DNB & 2 & 0.0168 & 2796 & -11 & 0.9012 \\ 
EBS & 26 &\textbf{0.0196} &\textbf{4308} &\textbf{-3} & \textbf{1.1715} \\ 
HSBC & 0 & 0.0172 & 3074 & -11 & 0.9462 \\ 
ING & 8 & 0.0186 & 3784 & -9 & 1.0814 \\ 
ISP & 2 & 0.0177 & 3288 & -11 & 0.9913 \\ 
KBC & 7 & \textbf{0.0191} & 4034 &\textbf{-5} & 1.1265 \\ 
LLOY & 11 & 0.0186 & 3748 &\textbf{-5} & 1.0814 \\ 
LUKN & 0 & 0.0168 & 2828 & -15 & 0.9012 \\ 
NDA & 0 & 0.0161 & 2380 & -11 & 0.8110 \\ 
RBS & 11 & 0.0181 & 3530 & -7 & 1.0363 \\ 
SEB & \textbf{33} & 0.0186 & 3748 & -9 & 1.0814 \\ 
GLE & 14 & 0.0177 & 3264 & -9 & 0.9913 \\ 
SGKN & 8 & 0.0172 & 3046 & -9 & 0.9462 \\ 
STAN & 5 & 0.0164 & 2594 & -11 & 0.8561 \\ 
SWED & 9 & 0.0172 & 3022 & -11 & 0.9462 \\ 
UCG & 0 &\textbf{0.0196} & \textbf{4264} & \textbf{-5} &\textbf{1.1715} \\ 
\hline
\end{tabular}
}
\label{T5}
\end{table}

\begin{table}[b]
\caption{Centrality Scores for Banks: financial crisis.}
\centering
\scalebox{0.99}{
\begin{tabular}{cccccc}
\hline
& Betweeness & Closeness & Laplacian & Centroid & LeaderRank \\ 
\hline
BMPS & 10 & {\textbf{0.0202}} & {\textbf{4510}} & {\textbf{-3}} & {\textbf{1.2382}} \\ 
BPM & 1 & 0.0158 & 2162 & -13 & 0.7796 \\ 
BBVA & 2 & 0.0169 & 2804 & -13 & 0.9172 \\ 
SAB & 2 & 0.0165 & 2606 & -11 & 0.8713 \\ 
BPES & 3 & 0.0177 & 3256 & -13 & 1.0089 \\ 
SAN & 1 & 0.0182 & 3434 & -11 & 1.0547 \\ 
BCV & 0 & 0.0161 & 2384 & -15 & 0.8254 \\ 
BARC & 11 & 0.0169 & 2760 & -13 & 0.9172 \\ 
BSKP & 0 & 0.0154 & 1968 & -15 & 0.7337 \\ 
BNP & 7 & 0.0197 & 4216 & -5 & 1.1923 \\ 
CBK & {\textbf{38}} & 0.0177 & 3280 & -5 & 1.0089 \\ 
ACA & 15 & 0.0177 & 3236 & -9 & 1.0089 \\ 
CSG & 3 & 0.0169 & 2816 & -9 & 0.9172 \\ 
DANSKE & 31 & {\textbf{0.0202}} & {\textbf{4506}} & {\textbf{-3}} & {\textbf{1.2382}} \\ 
DBK & 16 & 0.0182 & 3494 & -7 & 1.0547 \\ 
DNB & 6 & 0.0182 & 3482 & -9 & 1.0547 \\ 
EBS & 11 & 0.0173 & 2986 & -7 & 0.9630 \\ 
HSBC & 3 & 0.0173 & 3066 & -7 & 0.9630 \\ 
ING & 20 &{\textbf{0.0202}} & {\textbf{4506}} & {\textbf{-3} }& {\textbf{1.2382}} \\ 
ISP & 5 & 0.0186 & 3764 & -5 & 1.1006 \\ 
KBC & 9 & 0.0173 & 3078 & -9 & 0.9630 \\ 
LLOY & {\textbf{32}} & 0.0182 & 3498 & -9 & 1.0547 \\ 
LUKN & 2 & 0.0182 & 3482 & -9 & 1.0547 \\ 
NDA & 2 & 0.0181 & 3494 & -5 & 1.0547 \\ 
RBS & {\textbf{36}} & 0.0182 & 3494 & -11 & 1.0547 \\ 
SEB & 7 & 0.0169 & 2808 & -11 & 0.9172 \\ 
GLE & 18 & 0.0173 & 3006 & -9 & 0.9630 \\ 
SGKN & 1 & 0.0177 & 3232 & -7 & 1.0089 \\ 
STAN & 0 & 0.0177 & 3268 & -11 & 1.0089 \\ 
SWED & 18 & 0.0165 & 2610 & -7 & 0.8713 \\ 
UCG & 6 & 0.0173 & 3046 & -9 & 0.9630 \\ 
\hline
\end{tabular}
}
\end{table}

\begin{table}[b]
\caption{Centrality Scores for Banks: sovereign crisis.}
\centering
\scalebox{0.99}{
\begin{tabular}{cccccc}
\hline
& Betweeness & Closeness & Laplacian & Centroid & LeaderRank \\ 
\hline
BMPS & {\textbf{27}} & {\textbf{0.0197}} & {\textbf{4284}} & {\textbf{-3}} &{\textbf{1.1681}} \\ 
BPM & {\textbf{25}} & {\textbf{0.0197}} & {\textbf{4296}} & -5 & {\textbf{1.1681}} \\ 
BBVA & 7 & 0.0169 & 2792 & -15 & 0.8986 \\ 
SAB & 2 & 0.0186 & 3756 & -7 & 1.0783 \\ 
BPES & 6 & 0.0181 & 3562 & -9 & 1.0333 \\ 
SAN & 3 & 0.0161 & 2384 & -15 & 0.8087 \\ 
BCV & 1 & 0.0181 & 3542 & -9 & 1.0333 \\ 
BARC & 12 & 0.0173 & 3082 & -11 & 0.9435 \\ 
BSKP & 3 & {\textbf{0.0202}} &{\textbf{4578}} & -9 & {\textbf{1.2130}} \\ 
BNP & 21 & 0.0186 & 3808 & -7 & 1.0783 \\ 
CBK & {\textbf{48}} & 0.0192 & 4066 & {\textbf{0}} & 1.1232 \\ 
ACA & 19 & 0.0182 & 3578 & -7 & 1.0333 \\ 
CSG & 18 & 0.0186 & 3764 & -7 & 1.0783 \\ 
DANSKE & 3 & 0.0165 & 2574 & -13 & 0.8536 \\ 
DBK & 1 & 0.0173 & 3046 & -11 & 0.9435 \\ 
DNB & 6 & 0.0191 & 4046 & {\textbf{-3}} & 1.1232 \\ 
EBS & 21 & 0.0186 & 3808 & -7 & 1.0783 \\ 
HSBC & 0 & 0.0177 & 3292 & -11 & 0.9884 \\ 
ING & 1 & 0.0181 & 3562 & -7 & 1.0333 \\ 
ISP & 3 & 0.0181 & 3586 & -15 & 1.0333 \\ 
KBC & 3 & 0.0165 & 2606 & -15 & 0.8536 \\ 
LLOY & 20 & 0.0173 & 3074 & -9 & 0.9435 \\ 
LUKN & 1 & 0.0191 & 4070 & -9 & 1.1232 \\ 
NDA & 2 & 0.0165 & 2598 & -11 & 0.8536 \\ 
RBS & 12 & 0.0169 & 2880 & -11 & 0.8985 \\ 
SEB & 1 & 0.0161 & 2352 & -15 & 0.8087 \\ 
GLE & 11 & 0.0191 & 4010 & -9 & 1.1232 \\ 
SGKN & 2 & 0.0165 & 2594 & -13 & 0.8536 \\ 
STAN & 0 & 0.0157 & 2174 & -17 & 0.7638 \\ 
SWED & 3 & 0.0191 & 4058 & {\textbf{-3}} & 1.1232 \\ 
UCG & 20 & 0.0173 & 3018 & -9 & 0.9435 \\ 
\hline
\end{tabular}
}
\end{table}

\begin{table}[b]
\caption{Centrality Scores for Banks: post-crisis.}
\centering
\scalebox{0.99}{
\begin{tabular}{cccccc}
\hline
& Betweeness & Closeness & Laplacian & Centroid & LeaderRank \\ 
\hline
BMPS & \textbf{73} & \textbf{0.0208} & \textbf{4908} & \textbf{0} & \textbf{1.2056} \\ 
BPM & 12 & \textbf{0.0208} & \textbf{4972} & -5 & \textbf{1.2056} \\ 
BBVA & 5 & 0.0196 & 4396 & -11 & 1.1194 \\ 
SAB & 6 & 0.0168 & 2940 & -9 & 0.8611 \\ 
BPES & \textbf{25} & \textbf{0.0202} & \textbf{4702} & -9 & 1.1625 \\ 
SAN & 3 & 0.0186 & 3876 & -9 & 1.0333 \\ 
BCV & 1 & 0.0202 & 4662 & \textbf{-3} & \textbf{1.1625} \\ 
BARC & 1 & 0.0165 & 2674 & -15 & 0.8181 \\ 
BSKP & 0 & 0.0177 & 3368 & -11 & 0.9472 \\ 
BNP & 1 & 0.0186 & 3852 & -9 & 1.0333 \\ 
CBK & 6 & 0.0161 & 2416 & -13 & 0.7750 \\ 
ACA & 3 & 0.0173 & 3206 & -11 & 0.9042 \\ 
CSG & 3 & 0.0165 & 2698 & -17 & 0.8181 \\ 
DANSKE & 2 & 0.0202 & 4650 & -7 & 1.1625 \\ 
DBK & \textbf{25} & 0.0186 & 3868 & -9 & 1.0333 \\ 
DNB & 0 & 0.0173 & 3146 & -13 & 0.9042 \\ 
EBS & 4 & 0.0173 & 3134 & -13 & 0.9042 \\ 
HSBC & 0 & 0.0186 & 3888 & -9 & 1.0333 \\ 
ING & 5 & 0.0196 & 4384 & -9 & 1.1194 \\ 
ISP & 3 & 0.0161 & 2472 & -17 & 0.7750 \\ 
KBC & 6 & 0.0168 & 2888 & -13 & 0.8611 \\ 
LLOY & 9 & 0.0177 & 3424 & -11 & 0.9472 \\ 
LUKN & 0 & 0.0191 & 4150 & -9 & 1.0764 \\ 
NDA & 1 & 0.0177 & 3364 & -11 & 0.9472 \\ 
RBS & 5 & 0.0181 & 3658 & -11 & 0.9903 \\ 
SEB & 0 & 0.0186 & 3876 & -7 & 1.0333 \\ 
GLE & 14 & 0.0186 & 3944 & -5 & 1.0333 \\ 
SGKN & 0 & 0.0177 & 3408 & -17 & 0.9472 \\ 
STAN & 13 & 0.0202 & \textbf{4702} & -5 & 1.1625 \\ 
SWED & 3 & 0.0173 & 3174 & -7 & 0.9042 \\ 
UCG & \textbf{43} & 0.0197 & 4404 &\textbf{ -3} & 1.1194 \\ 
\hline
\end{tabular}
}
\end{table}

\begin{table}[b]
\caption{Centrality Scores for Countries: pre-crisis.}
\centering
\scalebox{0.99}{
\begin{tabular}{cccccc}
\hline
& Betweeness & Closeness & Laplacian & Centroid & LeaderRank \\ 
\hline
Italy & 4 & 0.0544 & 642 & -4 & 1.1000 \\ 
Spain & 1 & 0.0471 & 450 & -6 & 0.9000 \\ 
Switzerland &{\textbf{5}} & 0.0544 & 650 & -4 & 1.1000 \\ 
United Kingdom & {\textbf{6}} & {\textbf{0.0589}} & \textbf{740} & \textbf{0} & \textbf{1.2000} \\ 
France & 2 & \textbf{0.0589} & \textbf{744} & \textbf{-2} & \textbf{1.2000} \\ 
Germany & 1 & 0.0442 & 360 & -4 & 0.8000 \\ 
Denmark & 1 & 0.0505 & 544 & {\color{black}-2} & 1.0000 \\ 
Norway & 2 & 0.0416 & 274 & -8 & 0.7000 \\ 
Austria & 0 & 0.0416 & 294 & -6 & 0.7000 \\ 
Netherlands & 2 & 0.0544 & 638 & \textbf{-2} & 1.1000 \\ 
Belgium & \textbf{11} & 0.0544 & 642 & -4 & 1.1000 \\ 
Sweden & 1 & 0.0544 & 646 & -4 & 1.1000 \\ 
\hline
\end{tabular}
}
\end{table}

\begin{table}[b]
\caption{Centrality Scores for Countries: financial crisis.}
\centering
\scalebox{0.99}{
\begin{tabular}{cccccc}
\hline
& Betweeness & Closeness & Laplacian & Centroid & LeaderRank \\ 
\hline
Italy & \textbf{5} & 0.0473 & 434 & \textbf{-2} & 0.9000 \\ 
Spain & 0 & 0.0544 & 642 & -4 & 1.1000 \\ 
Switzerland & 0 & 0.0472 & 450 & -4 & 0.9000 \\ 
United Kingdom & 1 & 0.0506 & 548 & \textbf{-2} & 1.0000 \\ 
France & 3 & 0.0507 & 532 & -4 & 1.0000 \\ 
Germany & 2 & 0.0507 & 552 & -4 & 1.0000 \\ 
Denmark & \textbf{6} & 0.0545 & 634 & -6 & 1.1000 \\ 
Norway & 3 & \textbf{0.0590} & \textbf{748} & \textbf{-2} & \textbf{1.2000} \\ 
Austria & 3 & 0.0506 & 524 & -4 & 1.0000 \\ 
Netherlands & 2 & 0.0394 & 200 & -8 & 0.6000 \\ 
Belgium & \textbf{11} & \textbf{0.0590} & \textbf{748} & \textbf{-2} & \textbf{1.2000} \\ 
Sweden & 0 & 0.0506 & 548 & \textbf{-2} & 1.0000 \\ 
\hline
\end{tabular}
}
\end{table}

\begin{table}[b]
\caption{Centrality Scores for Countries: sovereign crisis.}
\centering
\scalebox{0.99}{
\begin{tabular}{cccccc}
\hline
& Betweeness & Closeness & Laplacian & Centroid & LeaderRank \\ 
\hline
Italy & \textbf{4} & 0.0590 & 800 & \textbf{0} & 1.0435 \\ 
Spain & 0 & 0.0589 & 808 & -6 & 1.0435 \\ 
Switzerland & 0 & 0.0472 & 494 & -6 & 0.7826 \\ 
United Kingdom & 1 & \textbf{0.0643} & \textbf{918} & -2 & \textbf{1.1304} \\ 
France & 3 & 0.0590 & 804 & -2 & 1.0435 \\ 
Germany & \textbf{7} & \textbf{0.0644} & \textbf{918} & \textbf{0} & \textbf{1.1304} \\ 
Denmark & 0 & 0.0544 & 694 & -4 & 0.9565 \\ 
Norway & 0 & 0.0544 & 694 & -4 & 0.9565 \\ 
Austria & 2 & 0.0506 & 592 & -4 & 0.8696 \\ 
Netherlands & 0 & 0.0545 & 698 & -4 & 0.9565 \\ 
Belgium & 1 & 0.0590 & 796 & -6 & 1.0435 \\ 
Sweden & 0 & 0.0590 & 796 & -6 & 1.0435 \\ 
\hline
\end{tabular}
}
\end{table}

\begin{table}[b]
\caption{Centrality Scores for Countries: post-crisis.}
\centering
\scalebox{0.99}{
\begin{tabular}{cccccc}
\hline
& Betweeness & Closeness & Laplacian & Centroid & LeaderRank \\ 
\hline
Italy & \textbf{6} & \textbf{0.0643} & \textbf{910} & \textbf{-2} & \textbf{1.1471} \\ 
Spain & 2 & 0.0544 & 682 & -4 & 0.9706 \\ 
Switzerland & 1 & 0.0544 & 686 & -4 & 0.9706 \\ 
United Kingdom & 3 & \textbf{0.0643} & \textbf{910} & \textbf{-2} & \textbf{1.1471} \\ 
France & \textbf{4} & 0.0590 & 792 & \textbf{-2} & 1.0588 \\ 
Germany & \textbf{4} & 0.0506 & 580 & \textbf{-2} & 0.8824 \\ 
Denmark & 0 & 0.0505 & 580 & -6 & 0.8824 \\ 
Norway & 0 & 0.0589 & 792 & -4 & 1.0588 \\ 
Austria & 0 & 0.0544 & 686 & -4 & 0.9706 \\ 
Netherlands & 0 & 0.0544 & 682 & -6 & 0.9706 \\ 
Belgium & 0 & 0.0544 & 686 & -6 & 0.9706 \\ 
Sweden & 0 & 0.0544 & 686 & -4 & 0.9706 \\ 
\hline
\end{tabular}
}
\label{T12}
\end{table}

The tables \ref{T5}-\ref{T12} show the centrality scores using the 5 aforementioned different measures. Banks/Countries that have a significant score are highlighted in bold values. The betweenness centrality measure seems to be the least correlated with the other 4 measures. This is likely due to the fact that it is the only measure that does not account the weights of edges in the financial network. 
For the bank network pre-crisis, Erste Group Bank and UniCredit are the more important banks within the financial network, followed by Banco Popular Espanol, Commerzbank AG and Credit Agricole. These banks are from different countries around Europe but are important financial institutions within their own country. During the financial crisis, the three critical banks are BMPS, Danske Bank and ING. This supports the fact that banks from Denmark and Netherlands were one of the worst affected banks during the financial crisis. This would have made them riskier within the financial network, resulting in higher centrality values, especially for measures that rely heavily on edge weights. \\
During the sovereign crisis, the banks that had the highest centrality scores were BMPS, Banca Popolare di Milano and Basler Kantonalbank. BMPS fell into severe financial trouble in 2012 due to increasing Italian government debt and lost more than a billion dollars, and was later involved in a scandal. The large risk associated with BMPS resulted in higher  CoRisk  values, which caused the bank to be a more central node in the financial network. BPM was also hit hard during the sovereign crisis and had large  CoRisk  values. These two banks continued to have difficulty after the sovereign crisis, and have the highest centrality scores for that period too. For the country network pre-crisis, the UK and France were the most important countries in the European financial network. When the financial crisis hit, Norway and Belgium became the more centrally located nodes in the network. Belgium was one of the European states most affected by the financial crisis, resulting in a higher  CoRisk  value between other countries. Norway, on the other hand, was not hugely affected by the financial crisis. The impact on Norway is therefore more likely to be due to its high connectedness with other European economies. Germany and the UK became the countries with the highest centrality scores during the sovereign crisis. These two countries are the largest economies in Europe, and the driving force of the European economy as a whole during the sovereign crisis. In Post-crises, Italy becomes the most central country in the network due to the failure of its banking system as explained above (BMPS and BPM being contributors to the overall risk). 

\subsubsection{Directed Acyclic Graph: Topological Sort}
A directed minimum spanning tree is also a directed acyclic graph, allowing us to do a causal analysis on the 8 graphs we had obtained earlier. The aim of this was to see which banks and countries transmitted the most risk (source nodes) to other nodes in the financial network. The key banks/countries for each period are shown below. These banks/countries are not necessarily the same as those with high centrality scores. Instead, they are financial institutions with a high importance in the European economy. For instance, Credit Agricole and Commerzbank represent the key roles played by France and Germany, which are considered as leaders of the European economy as a whole. 

Key Banks transmitted the most risks for each period are as follows,
\begin{itemize}
\item Pre-Crisis: Banque Cantonale Vaudoise, Barclays, Commerzbank, Credit Agricole,
\item Financial Crisis: Commerzbank, Credit Suisse, ING, RBS,
\item Sovereign Crisis: Banco Popular Espanol, Commerzbank, Credit Agricole, ING,
\item Post-crisis: BMPS, Banco Popular Espanol, Barclays, Commerzbank.
\end{itemize}
Key Countries transmitted the most risks for each period are as follows, 
\begin{itemize}
\item Pre-Crisis: France, Germany,
\item Financial Crisis: France, Germany,
\item Sovereign Crisis: France, Germany, 
\item Post-Crisis: Italy.
\end{itemize} 
\newpage
\subsubsection{Evolution of Country Financial Network}
MSTs are not the only method of visualising the results of our analysis. Since there are only 12 countries in our study, we could analyse the complete graph to see the evolution of risk transmission during the crisis periods. Below are four such graphs, where the size of the node represents the Laplacian Centrality score and the width of the edges represent the net CoRisk (sum of CoRisk$_{in}$ and CoRisk$_{out}$) between two countries. We can see that net CoRisk significantly increased from Pre-Crisis to Financial Crisis and slightly reduced during the Sovereign Crisis. However, it remained at a relatively higher level in the Post-Crisis period when compared with Pre-Crisis. The increased node sizes over time also revealed an interesting pattern about increased Laplacian Centrality (level of influence in the network). 
 \begin{figure}[H]
 \centering
\scalebox{0.6}{
  \begin{tabular}{@{}c@{}}
   \includegraphics[width=1.2\textwidth, height=12cm]{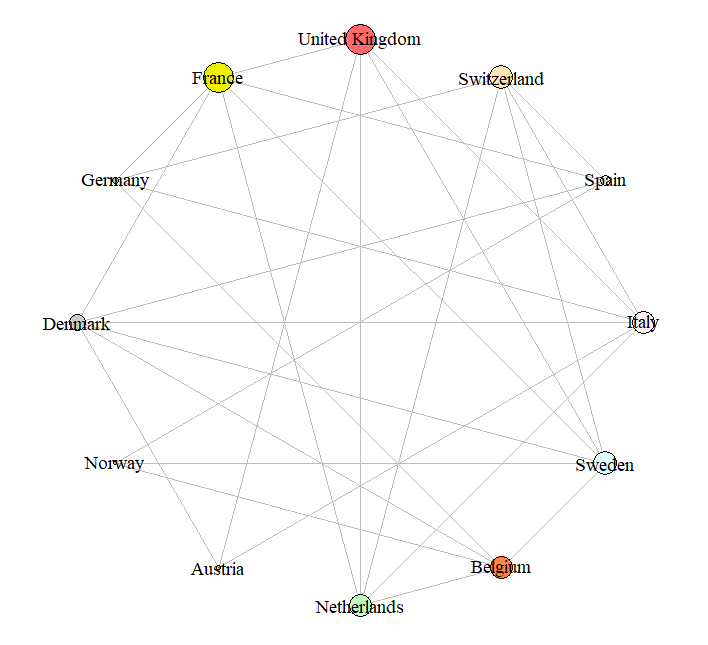} \\
   \includegraphics[width=1.2\textwidth, height=12cm]{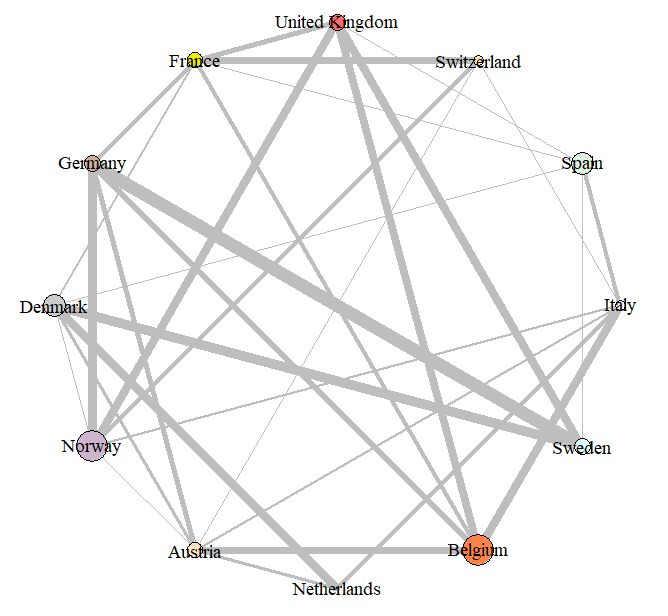} \\
\caption{Upper figure shows Pre-Crisis and  bottom plot report network in Financial Crisis. }
  \end{tabular}}
\label{PreCrisisGraph}
\end{figure}
 \begin{figure}[H]
 \centering
\scalebox{0.72}{
  \begin{tabular}{@{}c@{}}
    \includegraphics[width=1.2\textwidth, height=12cm]{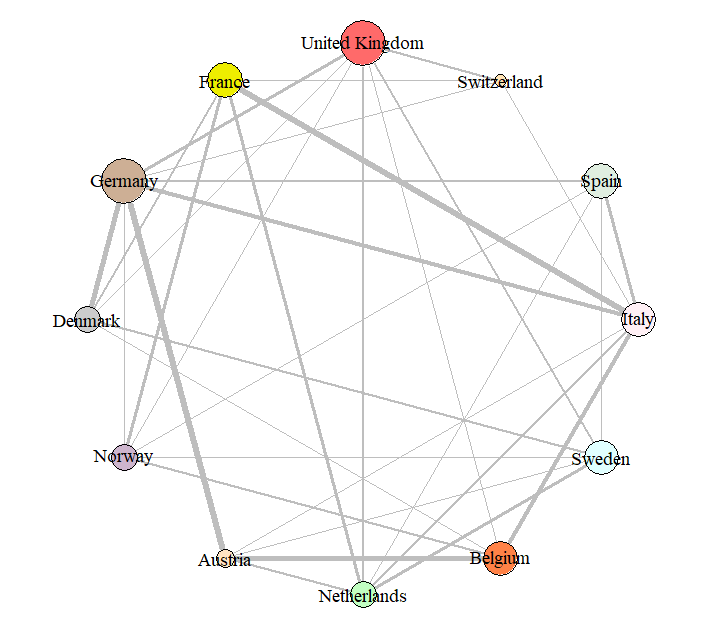}\\
\includegraphics[width=1.2\textwidth, height=12cm]{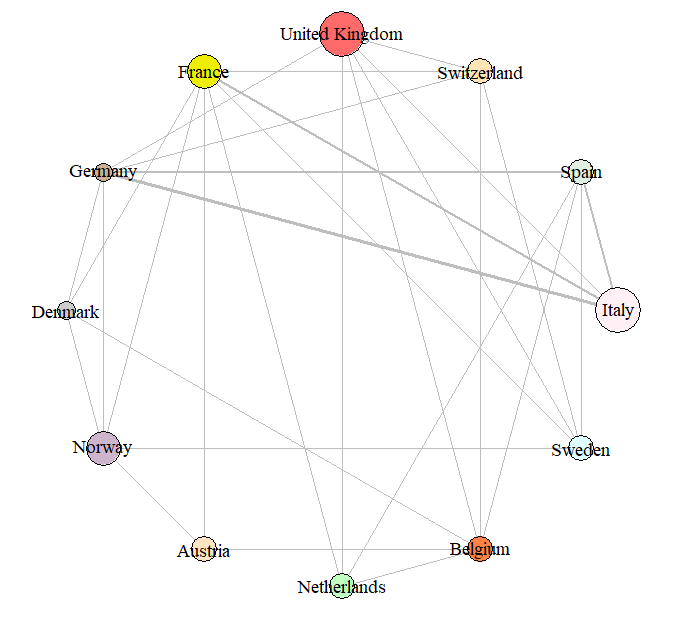} \\
\caption{Upper figure shows Sovereign Crisis, while the bottom plot shows network in Post-Crisis. }
  \end{tabular}}
\label{PostCrisisGraph}
\end{figure}

\subsection{Latent position model empirical results}
\begin{figure}[htb]
 \centering
 \includegraphics[width = 0.495\textwidth]{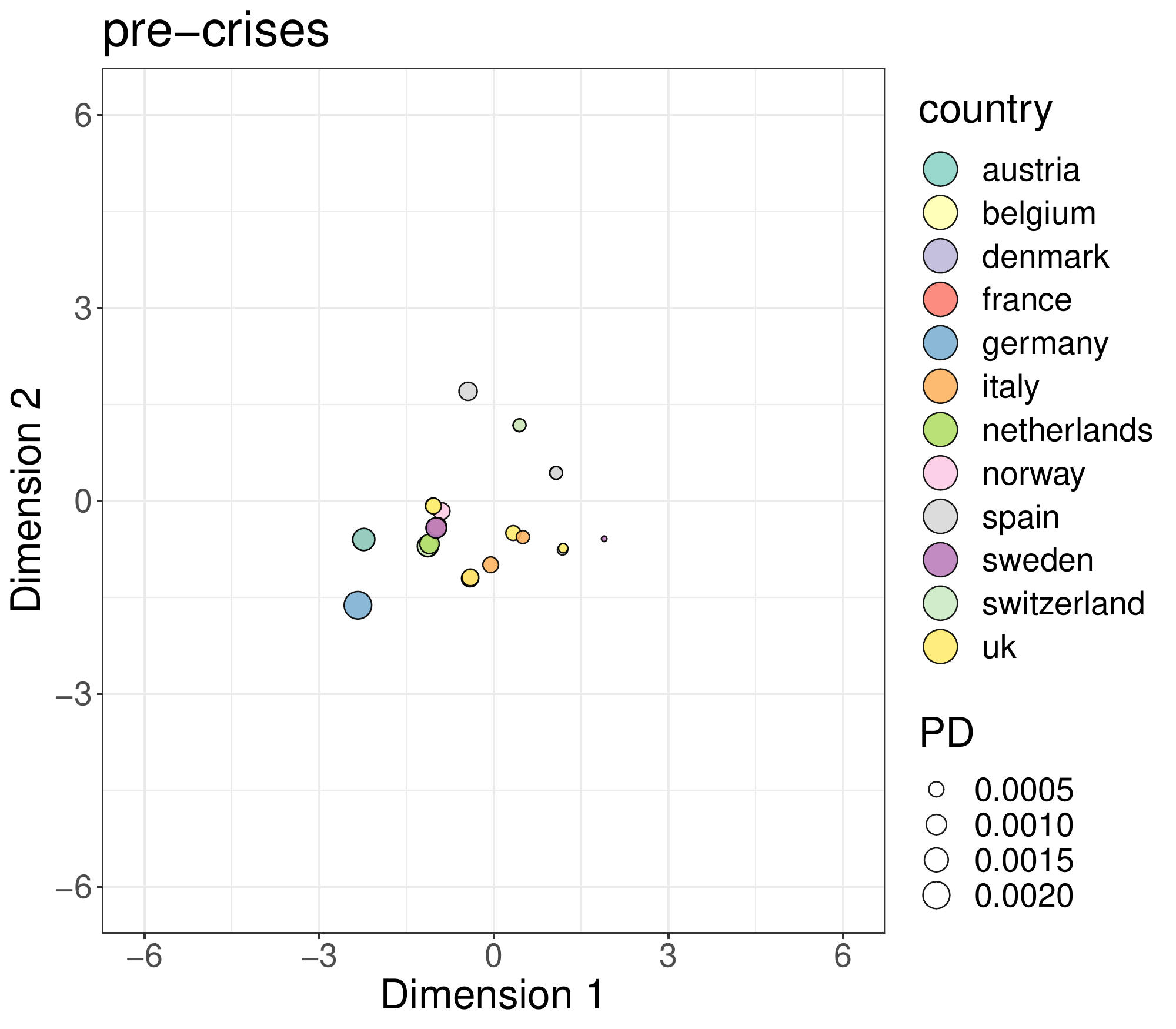}
 \includegraphics[width = 0.495\textwidth]{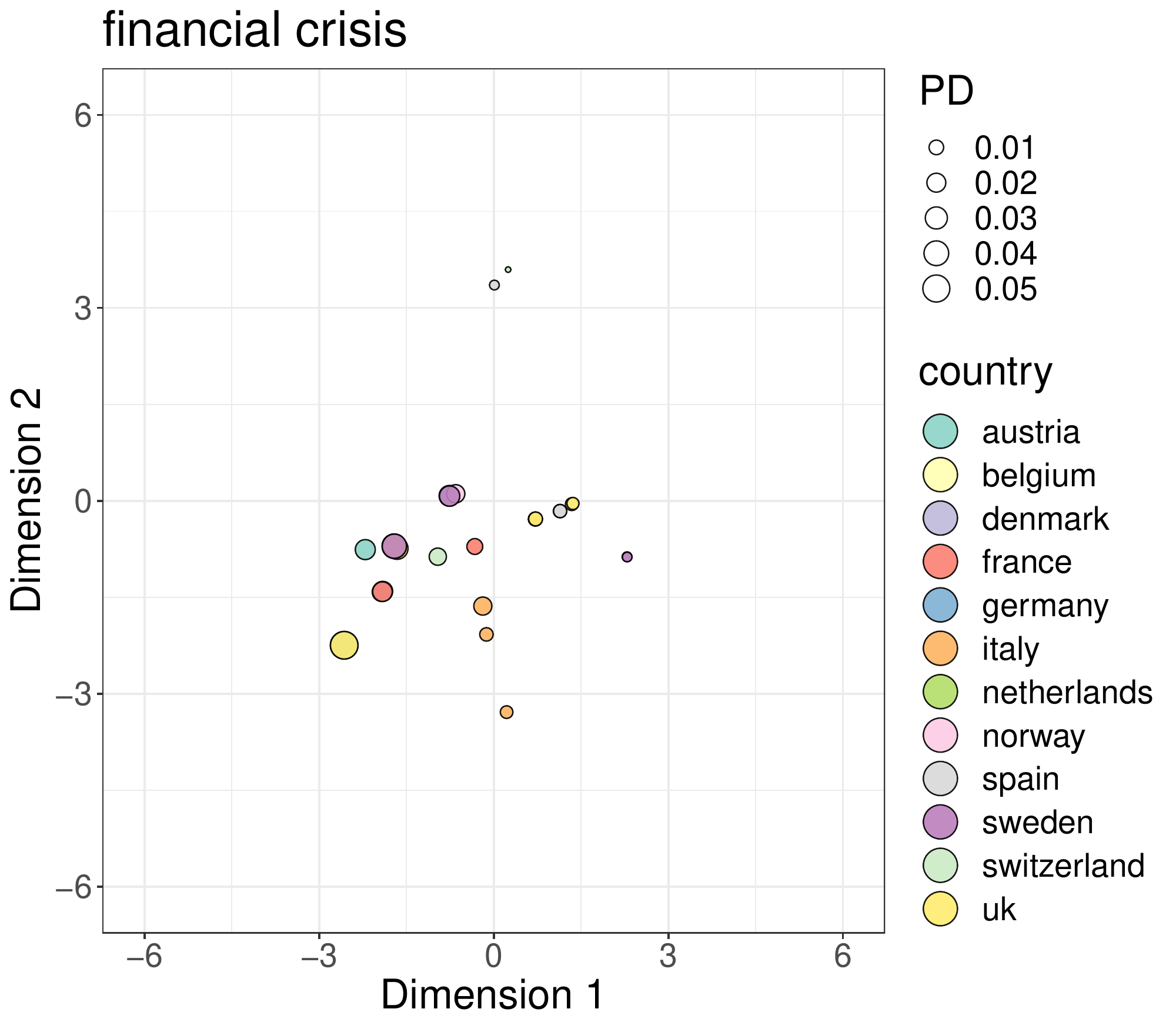}\\
 \includegraphics[width = 0.495\textwidth]{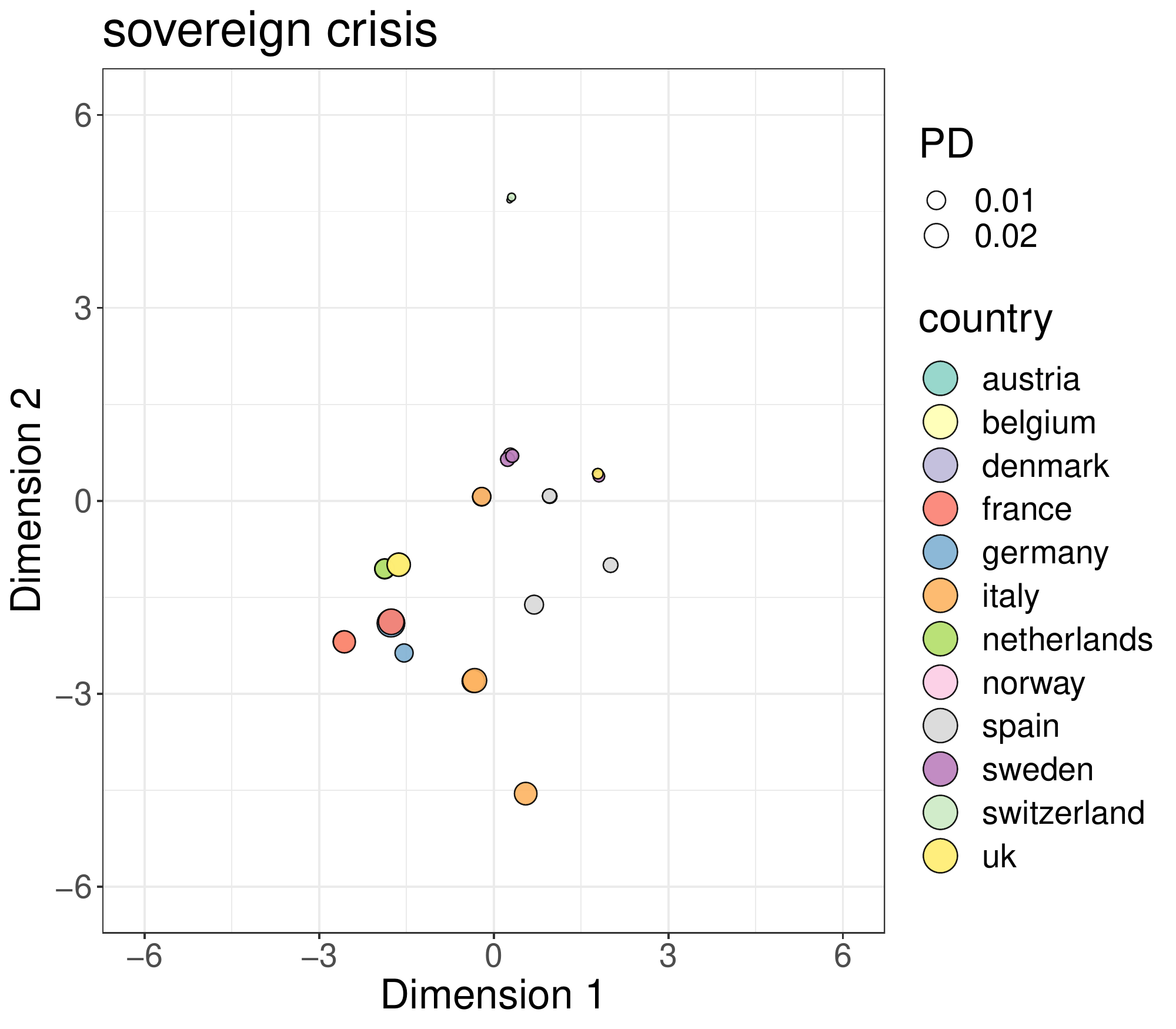}
 \includegraphics[width = 0.495\textwidth]{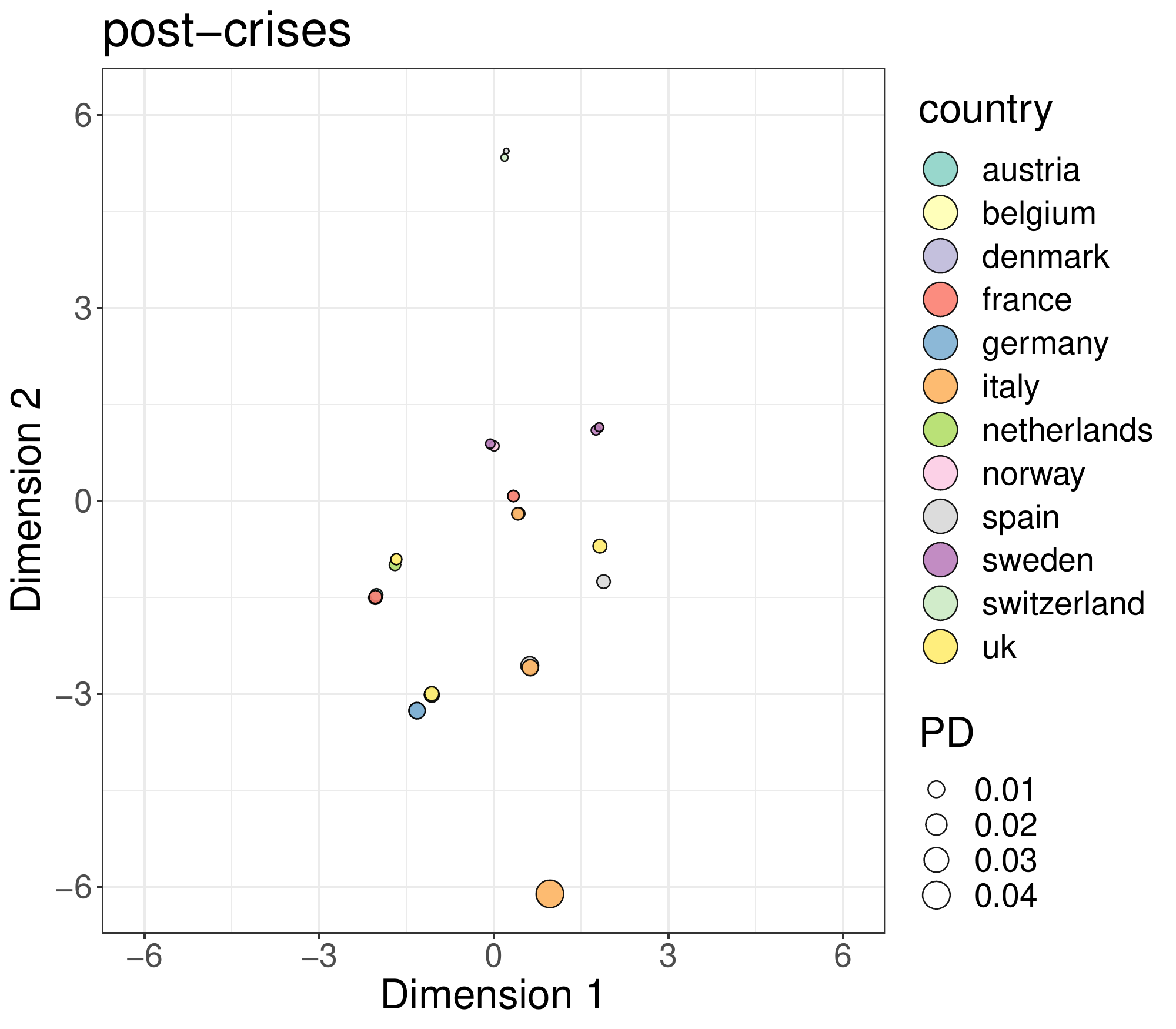}
 \caption{Estimated Latent Position Model for each of the time periods, with color of the nodes representing their country. Note that the axes have the same range for all panels. For visualisation purposes, Basler Kantonalbank and Luzerner Kantonalbank are excluded since they are located far from the rest of the banks.}
 \label{fig:lpm_countries}
\end{figure}

\begin{figure}[!htb]
 \centering
 \includegraphics[width = 0.495\textwidth]{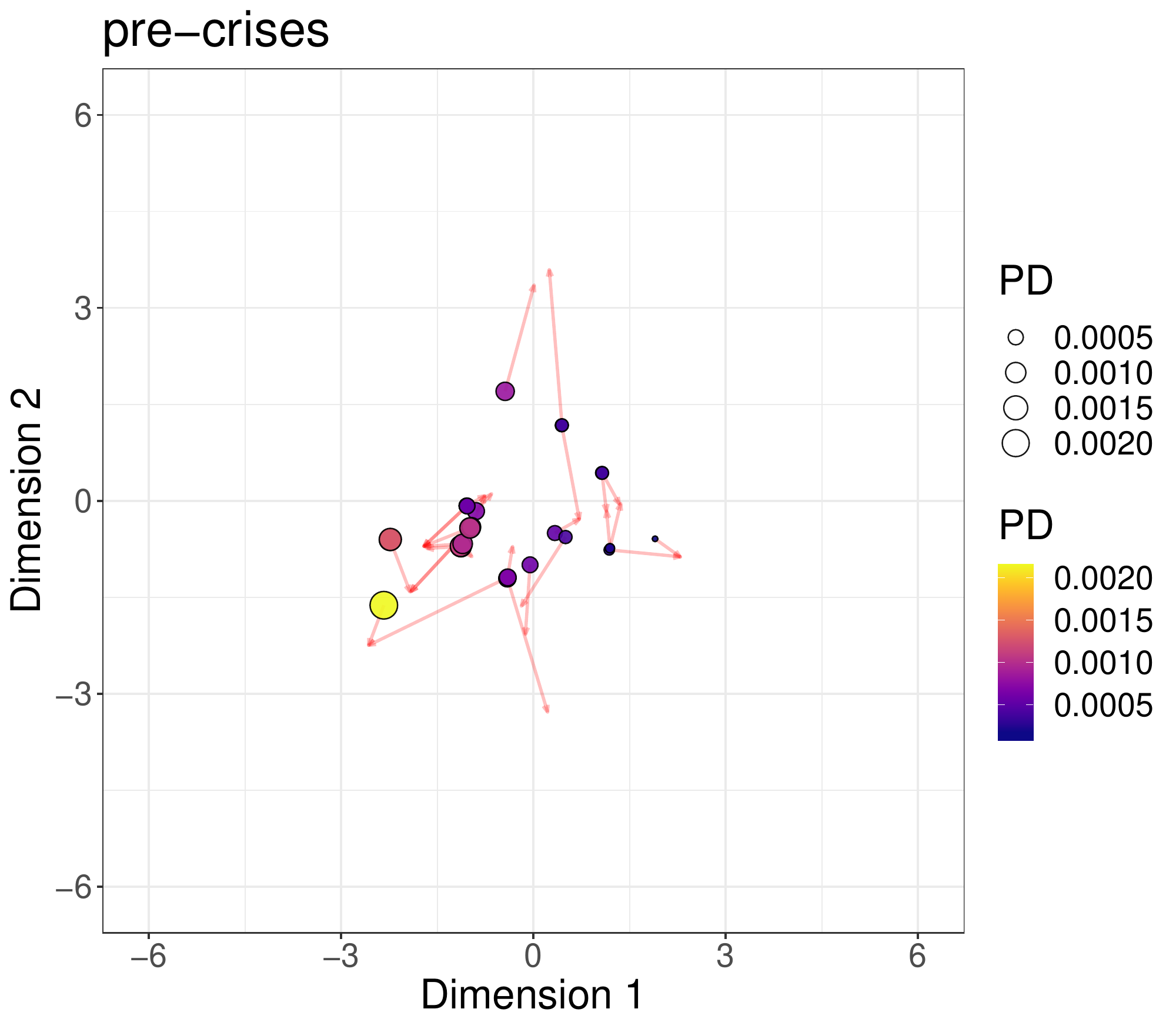}
 \includegraphics[width = 0.495\textwidth]{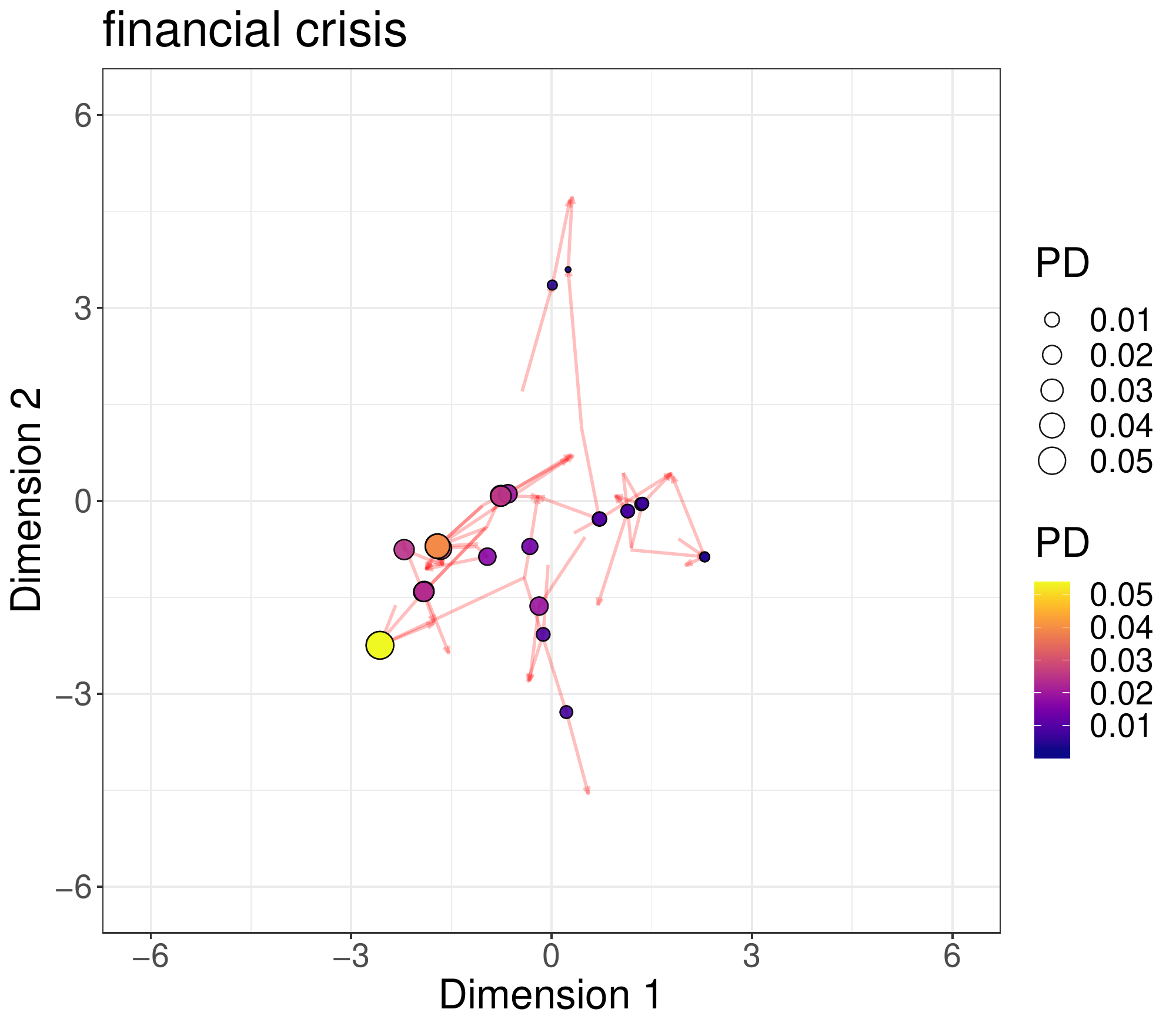}\\
 \includegraphics[width = 0.495\textwidth]{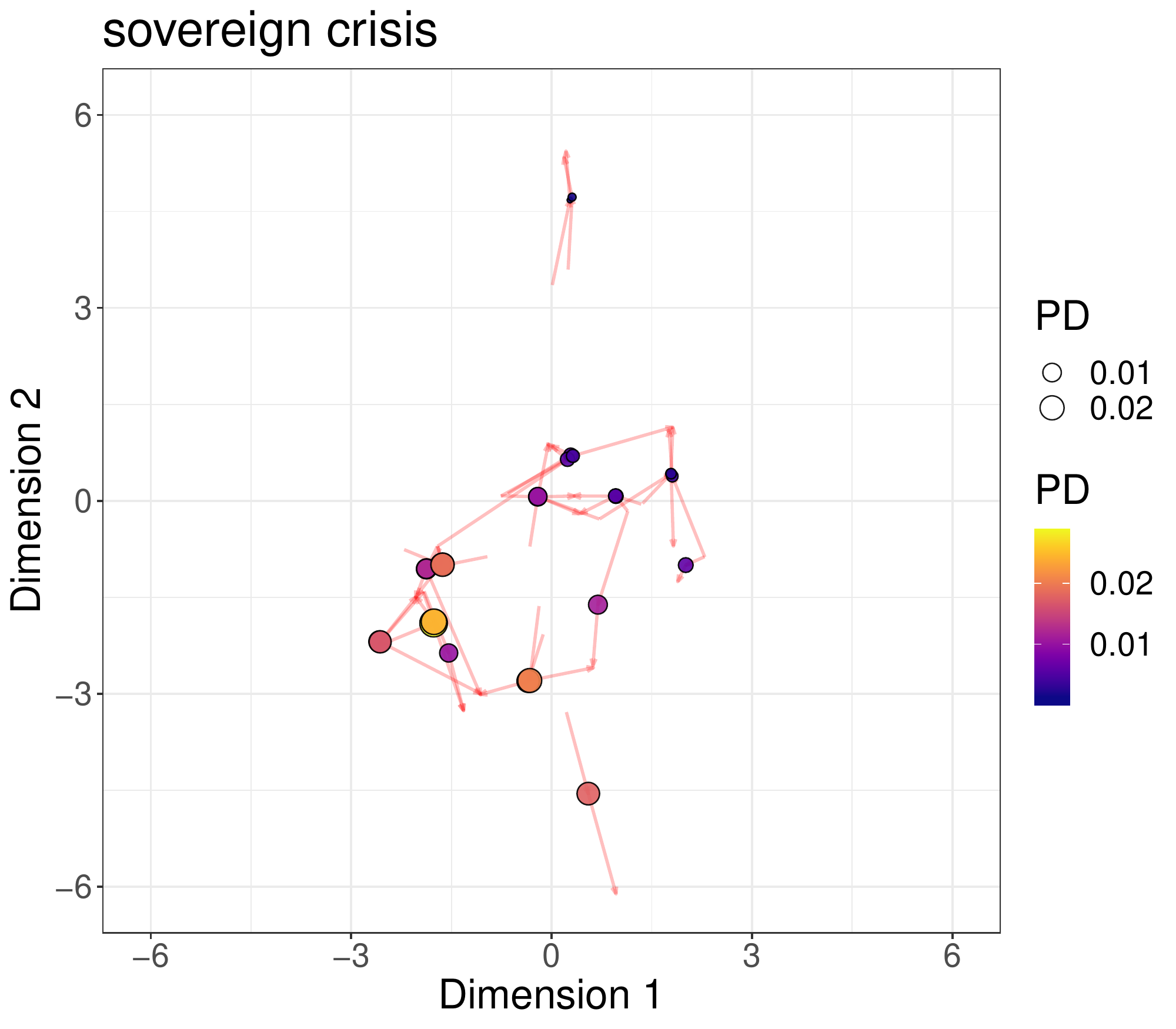}
 \includegraphics[width = 0.495\textwidth]{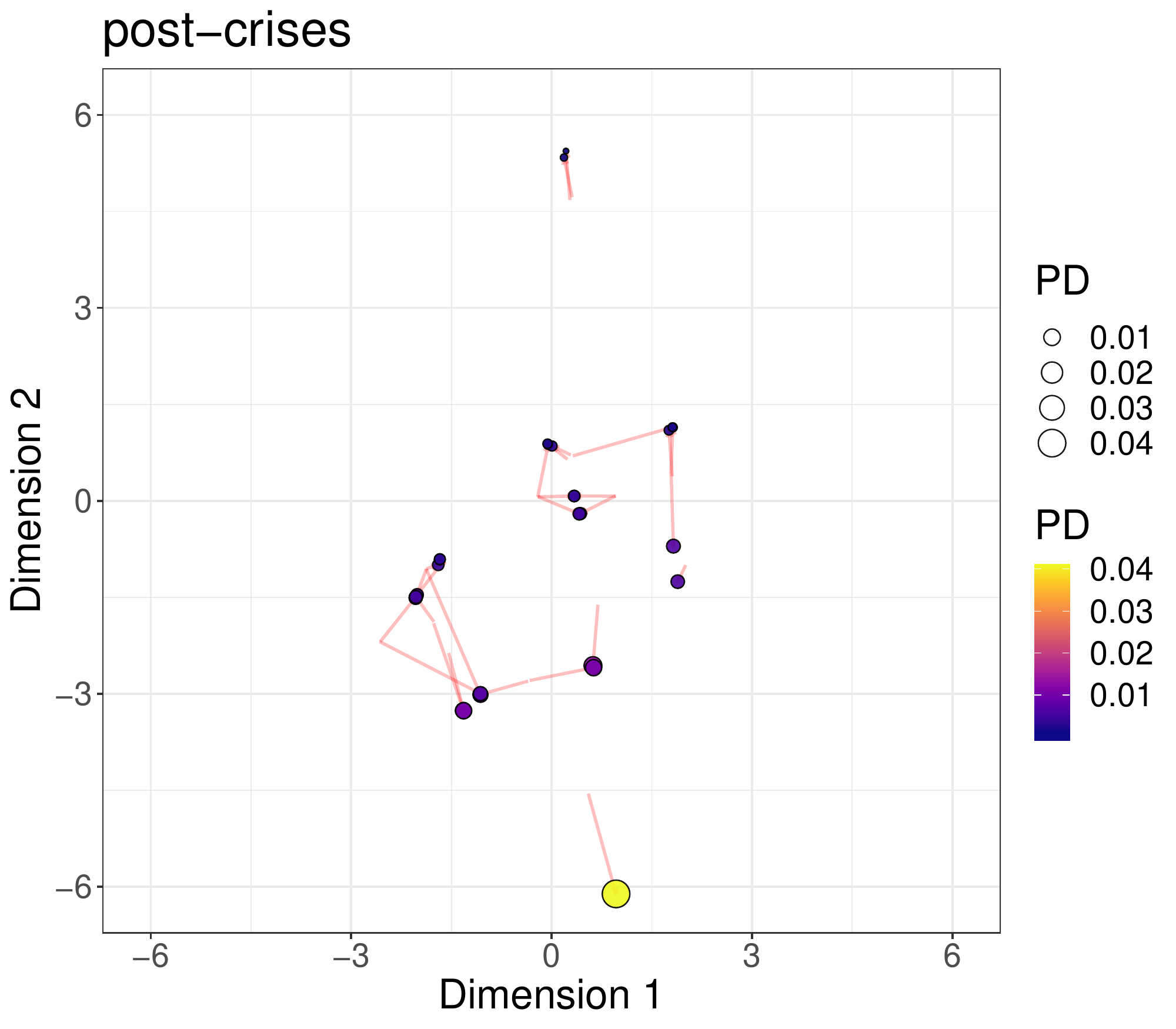}
 \caption{Estimated Latent Position Model for each of the time periods, with color of the nodes representing the associated default probability.
 The arrows represent the movement of each of the nodes from the previous time frame into the next. For visualisation purposes, Basler Kantonalbank and Luzerner Kantonalbank are excluded since they are located far from the rest of the banks.}
 \label{fig:lpm_pd}
\end{figure}

The main information that we obtain from our fitted LPM model are the latent positions of all financial institution in each of the periods.
We show these latent spaces in Figure \ref{fig:lpm_countries} and Figure \ref{fig:lpm_pd}, by comparing them with the corresponding country and probability of default, respectively.

Since not all the available information can be shown in these plots, we also include, as additional materials, an interactive version of the same plots which implements more features and visualisation tools. In Figure \ref{fig:lpm_countries}, the countries and probabilities of default of the financial institutions are highlighted within the latent space.
Before the crises, the points tend to be close to each other, signaling a dense status of the system, whereby risk may be easily spread.
We note that two Swiss banks are not shown in the plot since they are located far from the other institutions, and that they also show a low default probability in all time periods.

Starting from the second time period (which corresponds to the 2008 financial crisis), we observe that the latent visualisation expands and creates clusters in the space. The expansion can be interpreted as a measure to counteract risk from partners, and reduce the overall correlation.
Regarding the presence of clusters, our interpretation is rather straightforward: clustered institutions tend to be close to each other and will tend to be more contagious towards one another. As a consequence, they are more likely to exhibit similar default probabilities. 
By contrast, we expect less contagiousness between clusters, especially if they are located far apart.

\begin{remark}
It is important to note that these clusters highlight some associations between banks that may play a crucial role within the financial system.
These associations can help in identifying key institutions that can pose systemic risk concerns, but, also, they may help in understanding the dynamics of the spread of debt.
\end{remark}
We note a tendency to create assortative behaviour with respect to the countries, intended as the fact that banks of the same country are close to each other and thus share higher risks.
The Spanish, Swedish, and Italian banks (separately) seem to exhibit very similar behaviour throughout the period.
The Swiss banks are located in the outskirts of the latent space and seem to be not particularly influenced by the other institutions, throughout the study.
Many associations that are exhibited in these plots are not related to countries: an example could be the association between BARC and SWED during the financial crisis, or between ING and LLOY during the sovereign crisis.

In Figure \ref{fig:lpm_pd}, we highlight the dynamic nature of our model.
Both the colors and size of the points describe the default probability associated to the bank.
Also, we include two oriented segments to highlight the position changes with respect to the previous and following time frames.
This plot highlights that the banks with a higher default probability tend to be located near the centre and the lower regions of the space.
The model provides an interesting view on the dynamics for the bank BMPS, which starts from a central position, but then drifts away from all other banks while increasing its probability of default.
This is clearly in agreement with the idiosyncratic nature of the financial distress experienced by this bank during the period.

\subsubsection{A measure of systemic risk derived from the LPM}
\begin{figure}[!htb]
 \centering
 \includegraphics[width = 0.495\textwidth]{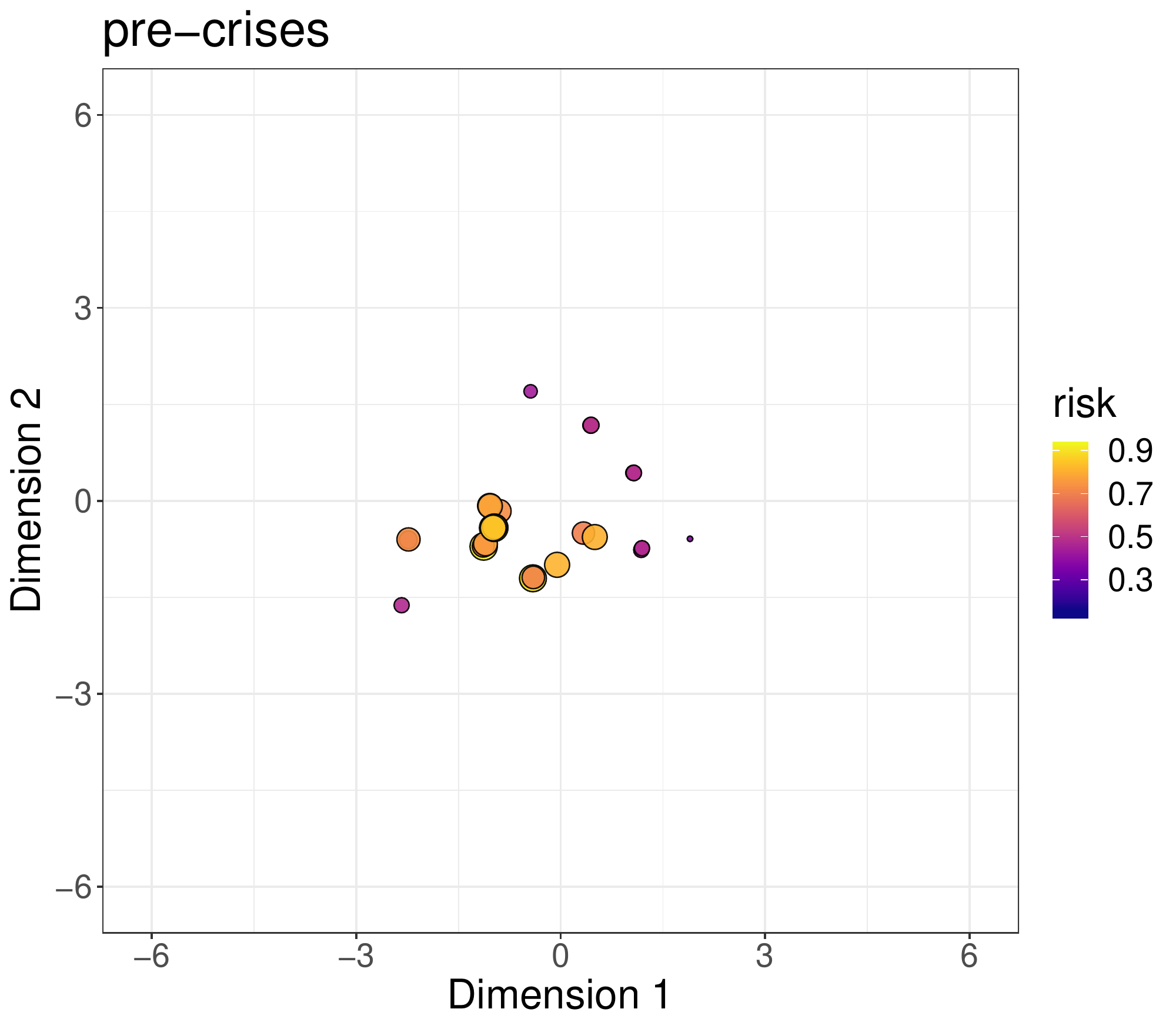}
 \includegraphics[width = 0.495\textwidth]{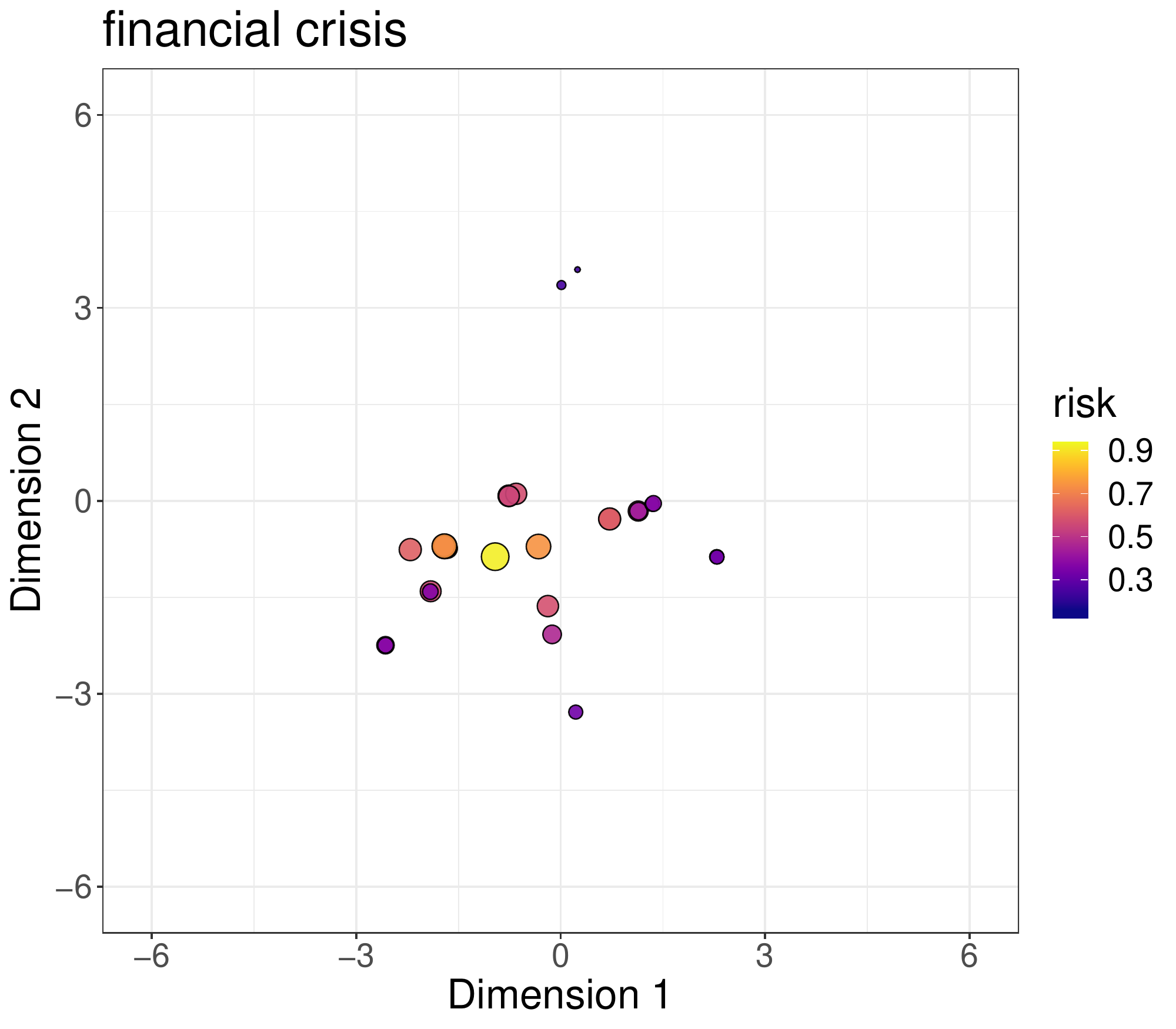}\\
 \includegraphics[width = 0.495\textwidth]{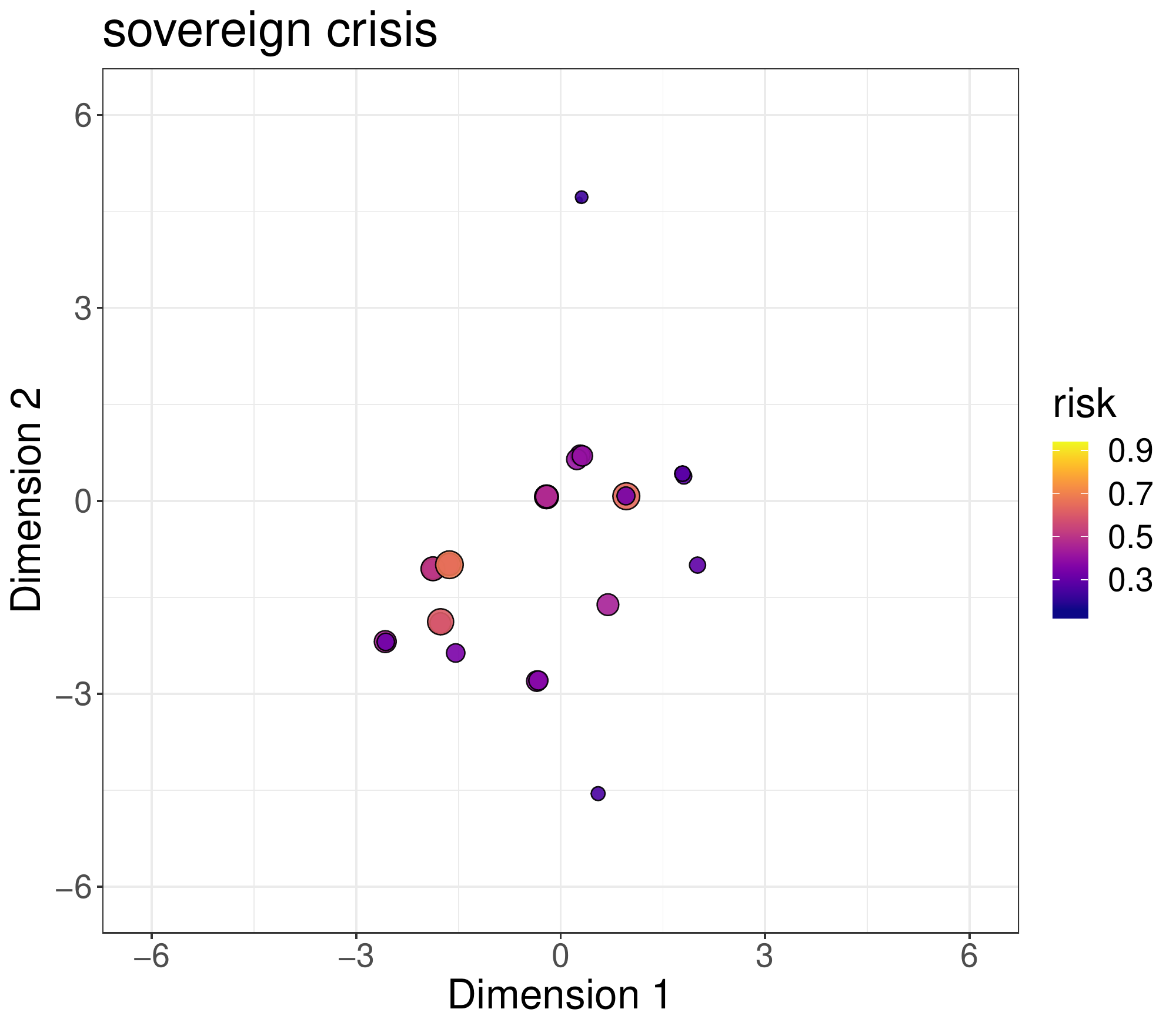}
 \includegraphics[width = 0.495\textwidth]{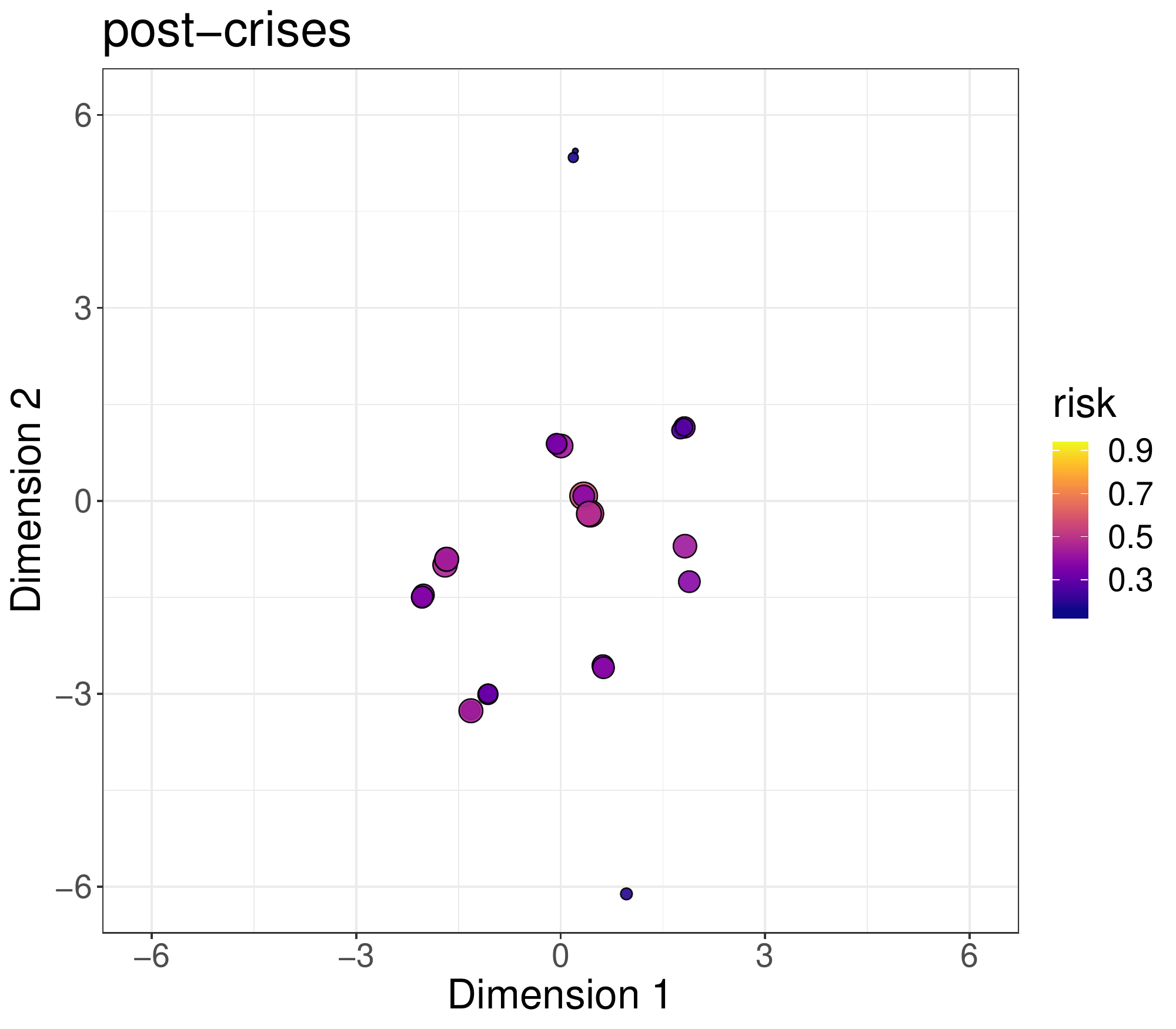}
 \caption{Estimated LPM for each of the time periods, with color of the nodes representing the estimated risk received from the system. Note that the axes have the same range for all panels. For visualisation purposes, Basler Kantonalbank and Luzerner Kantonalbank are excluded since they are located far from the rest of the banks.}
 \label{fig:lpm_risk}
\end{figure}

We use our inferred LPM to derive a measure of risk associated to each of the financial institutions.
Since we are interested in the part of risk that is received from the system, we can use the respective latent positions of all pairs of institutions to derive an interpretable quantity.
In the LPM view, the two necessary conditions for two institutions to affect each other are:
\begin{enumerate}
 \item the institutions are joined with an edge;
 \item the institutions are close in the latent space.
\end{enumerate}
As a consequence, we propose to calculate, for each institution, the inverse of the average Euclidean distance from its neighbours.
This index will be very high when the neighbours are on average very close, signalling that the banks will affect each other to a high degree.
By contrast, a low value of this index will indicate that the neighbours are not located close to the bank, so they cannot have much of an influence on it. 
Figure \ref{fig:lpm_risk} shows the banks in all time periods highlighting their measured systemic risk level.
Clearly, banks at higher risk are those located close to the centre of the space, and those that have several other banks nearby.
The institutions located in the outskirts of the space exhibit low risk.
We refer the reader to the supplementary materials for more information regarding specific banks.

\begin{figure}[htb]
 \centering
 \includegraphics[width = 0.7\textwidth]{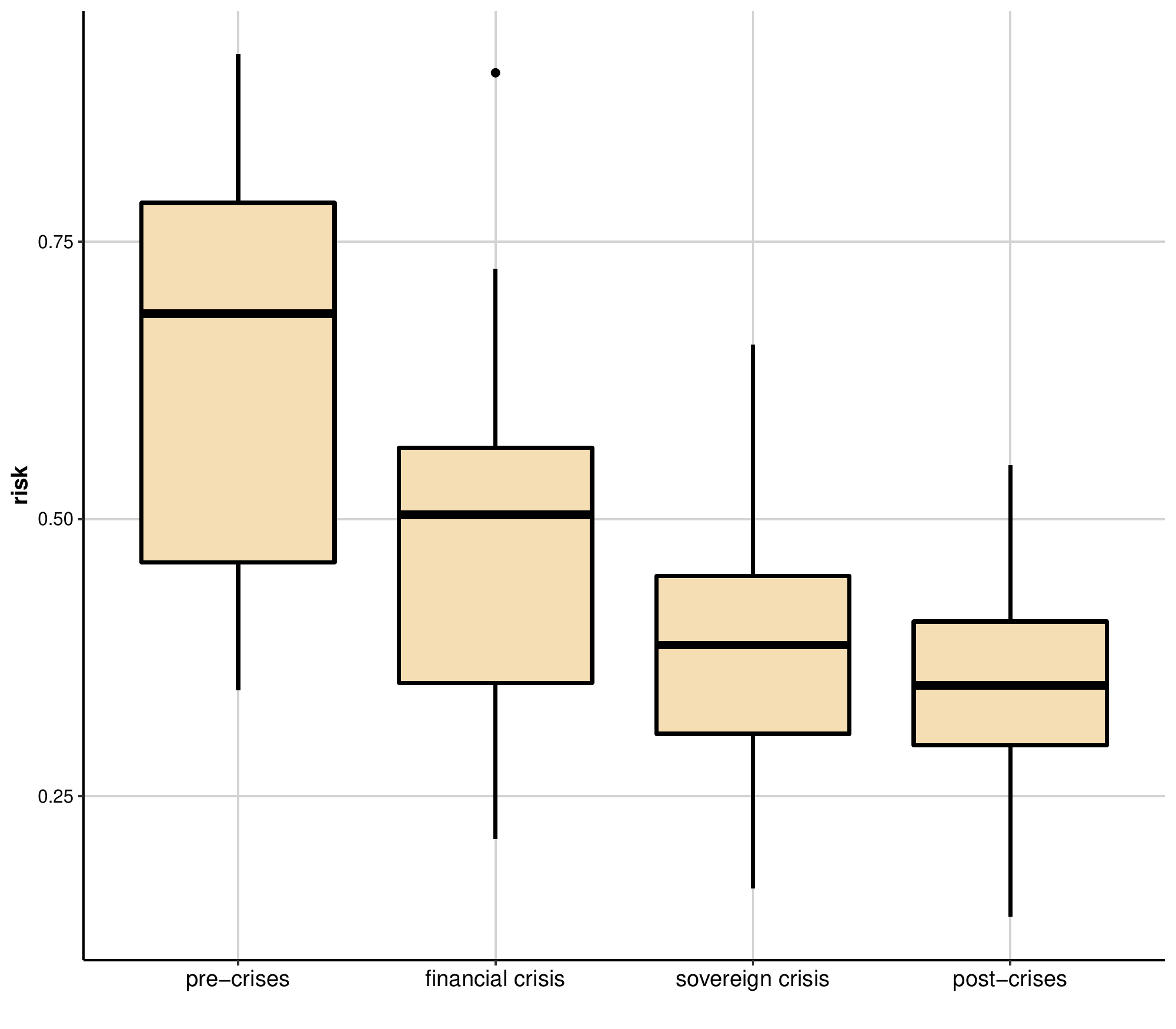}
 \caption{The boxplots represent the distribution of the measure of systemic risk across all the financial institutions in each of the time frames.}
 \label{fig:lpm_risk_global}
\end{figure}

From a more global perspective, we summarise the same information using boxplots in Figure \ref{fig:lpm_risk_global}.
This figure shows the distribution of the risk measure across time, signalling a clear decrease in systemic risk over time.
This can be interpreted as the model signalling a general tendency towards better financial stability.
\newpage
\section{Conclusions}\label{sec:conclusions}
The recent crises have changed drastically the structure of the European financial system. While the 2008 financial crisis has hit globally, the sovereign crisis has had a more lasting impact on certain European banks.
We have adopted a number of descriptive measures and model-based measures to obtain a new and different perspective on the European financial system.
The $CoRisk$ measure allowed us to see which banks and countries were more systemically important within the financial network during the financial and sovereign crises. In particular, we have disentangled the different driving forces of risk using the connectedness and the default probabilities. The combination of the visualisation tools and of the centrality measures were able to identify the main sources of the risk.
Regarding our latent space model, the results show that the banks expanded in the latent space during the crisis, presumably responding to the onset of the 2008 financial crisis. The model signals a general trend towards financial stability whereby the risk from other institutions remains contained.
In fact, after 2008, the latent point process exhibits a more sparse and clustered structure, which is known to be more resilient to targeted attacks or defaults.
The latent space approach also enabled us to isolate and measure the risk caused by the financial system, thus introducing a new index for systemic risk and providing additional evidence for the trend towards financial stability.\\
Regarding possible extensions of our work, we point out that the heuristic simulated annealing was used because of its speed and theoretical guarantees, however other approaches may be viable. In the Bayesian setting, one may design a sampler to obtain approximate draws from the posterior of the distribution of interest. This would also permit an assessment of the uncertainty associated to the positioning of the nodes in the latent space. 
In addition, our proposed solution can be implemented on a variety of financial decision-making platforms, enabling individual users to map complex financial systems and make better data-driven financial decisions.

\section{Appendix}
\subsection{Proof of Proposition \ref{prop:gaussian_1}}\label{app:gaussian_1}
\begin{proof}
We wish to study the density $\pi\left( y_i^{(t)} \middle\vert \textbf{y}_{-(i,t)}, \textbf{Z}, \textbf{X} \right)$, up to a proportionality constant that does not depend on $y_i^{(t)}$:
\begin{equation}
 \begin{split}
  \pi\left( y_i^{(t)} \middle\vert \textbf{y}_{-(i,t)}, \textbf{Z}, \textbf{X} \right)
  &= \frac{ \pi\left( \textbf{y} \middle\vert \textbf{Z}, \textbf{X} \right) }{ \pi\left( \textbf{y}_{-(i,t)} \middle\vert \textbf{Z}, \textbf{X} \right) }
  \propto f\left( \textbf{y} \middle\vert \textbf{Z}, \textbf{X} \right) \\
  &\propto \exp\left\{ - \frac{1}{2}\sum_{t=1}^{T} \sum_{i=1}^N \sum_{\substack{j=1 \\ j\neq i}}^N x_{ij}^{(t)} \eta_{ij}^{(t)} \left( y_i^{(t)} - y_j^{(t)} \right)^2\right\} \\
  &\propto \exp\left\{ - \frac{1}{2} \sum_{\substack{j=1 \\ j\neq i}}^N x_{ij}^{(t)} \eta_{ij}^{(t)} \left( y_i^{(t)} - y_j^{(t)} \right)^2\right\}\\
  &\propto \exp\left\{ - \frac{1}{2}\left[  \mathcal{A}_{i}^{(t)} \left(y_i^{(t)}\right)^2 - 2\mathcal{B}_{i}^{(t)} y_i^{(t)} \right] \right\}
 \end{split}
\end{equation}
where
$$
\mathcal{A}_{i}^{(t)} = \sum_{\substack{j=1 \\ j\neq i}}^N x_{ij}^{(t)} \eta_{ij}^{(t)},
$$
$$
\mathcal{B}_{i}^{(t)} = \sum_{\substack{j=1 \\ j\neq i}}^N x_{ij}^{(t)} \eta_{ij}^{(t)} y_j^{(t)}.
$$
Then:
\begin{equation}
 \begin{split}
  \pi\left( y_i^{(t)} \middle\vert \textbf{y}_{-(i,t)}, \textbf{Z}, \textbf{X} \right)
  &\propto \exp\left\{ - \frac{1}{2}\left[  \mathcal{A}_{i}^{(t)} \left(y_i^{(t)}\right)^2 - 2\mathcal{B}_{i}^{(t)} y_i^{(t)} \right] \right\} \\
  &\propto \exp\left\{ - \frac{\mathcal{A}_{i}^{(t)}}{2}\left[ \left(y_i^{(t)}\right)^2 - 2\frac{\mathcal{B}_{i}^{(t)}}{\mathcal{A}_{i}^{(t)}} y_i^{(t)} + \left(\frac{\mathcal{B}_{i}^{(t)}}{\mathcal{A}_{i}^{(t)}}\right)^2   \right] \right\} \\
  &\propto \exp\left\{ - \frac{\mathcal{A}_{i}^{(t)}}{2}\left[ y_i^{(t)} - \frac{\mathcal{B}_{i}^{(t)}}{\mathcal{A}_{i}^{(t)}} \right]^2 \right\} \\
 \end{split}
\end{equation}
which is proportional to a Gaussian density with the following mean and variance:
$$
\mu_i^{(t)} = \frac{\mathcal{B}_{i}^{(t)}}{\mathcal{A}_{i}^{(t)}},
\quad \quad \quad
\nu_i^{(t)} = \frac{1}{\mathcal{A}_{i}^{(t)}}.
$$ 
\end{proof}

\subsection{Summary of Default Probabilities by Bank}
\begin{table}[H]
\caption{Summary of Default Probabilities by Bank.}\label{PDbyBank}
\begin{center}
\scalebox{0.79}{
\begin{tabular}{lcccccc}
\hline
& Mean & SD & Maximum & Minimum & Skewness & Kurtosis \\ 
\hline
Banca Monte dei Paschi di Siena (BMPS) & 1.83\% & 2.54\% & 18.93\% & 0.04\% & 2.46 & 7.39 \\ 
Banca Popolare di Milano (BPM) & 0.77\% & 0.88\% & 4.6\% & 0.03\% & 1.66 & 2.24 \\ 
Banco Bilbao Vizcaya (BBVA) & 0.36\% & 0.43\% & 3.29\% & 0.02\% & 2.79 & 10.06 \\ 
Banco de Sabadell (SAB) & 0.35\% & 0.31\% & 1.51\% & 0.01\% & 1.11 & 1 \\ 
Banco Popular Espanol S.A (BPES) & 0.79\% & 1.02\% & 4.65\% & 0\% & 1.89 & 2.73 \\ 
Banco Santander (SAN) & 0.36\% & 0.44\% & 3.16\% & 0.02\% & 2.56 & 7.75 \\ 
Banque Cantonale Vaudoise (BCV) & 0.1\% & 0.14\% & 0.74\% & 0.01\% & 2.38 & 5.23 \\ 
Barclays (BARC) & 1.04\% & 1.53\% & 16.97\% & 0.04\% & 3.56 & 17.68 \\ 
Basler Kantonalbank (BSKP) & 0.08\% & 0.15\% & 0.98\% & 0\% & 2.29 & 4.84 \\ 
BNP Paribas (BNP) & 0.57\% & 0.76\% & 5.41\% & 0.02\% & 2.69 & 8.24 \\ 
Commerzbank AG (CBK) & 2.03\% & 3.1\% & 23.37\% & 0.09\% & 3.03 & 10.98 \\ 
Credit Agricole (ACA) & 1.1\% & 1.41\% & 7.78\% & 0.04\% & 1.99 & 3.63 \\ 
Credit Suisse Group (CSG) & 0.83\% & 0.98\% & 4.91\% & 0.05\% & 1.77 & 2.7 \\ 
Danske Bank (DANSKE) & 0.78\% & 1.71\% & 16.21\% & 0.01\% & 4.47 & 24.12 \\ 
Deutsche Bank (DBK) & 1\% & 1.52\% & 10.73\% & 0.04\% & 3.14 & 11.22 \\ 
DNB ASA (DNB) & 0.54\% & 1.13\% & 9.11\% & 0.03\% & 3.93 & 15.93 \\ 
Erste Group Bank AG (EBS) & 0.93\% & 1.39\% & 9.4\% & 0.06\% & 2.73 & 8.07 \\ 
HSBC & 0.2\% & 0.33\% & 2.91\% & 0.01\% & 3.82 & 17.08 \\ 
ING group (ING) & 1.12\% & 2.15\% & 21.55\% & 0.04\% & 3.99 & 20.3 \\ 
Intesa Sanpaolo (ISP) & 0.53\% & 0.63\% & 3.59\% & 0.02\% & 1.63 & 2.06 \\ 
KBC Bancassurance Holding S.A. (KBC) & 1.17\% & 2.09\% & 14.73\% & 0.02\% & 2.8 & 8.27 \\ 
Lloyds Banking Group (LLOY) & 1.26\% & 2.09\% & 16.04\% & 0.03\% & 2.7 & 8.95 \\ 
Luzerner Kantonalbank (LUKN) & 0.01\% & 0.01\% & 0.06\% & 0\% & 2.6 & 7.39 \\ 
Nordea (NDA) & 0.17\% & 0.21\% & 1.51\% & 0\% & 2.57 & 9.26 \\ 
Royal Bank of Scotland (RBS) & 1.62\% & 2.69\% & 16.11\% & 0.02\% & 2.82 & 8.07 \\ 
SEB AB (SEB) & 0.66\% & 1.29\% & 9.62\% & 0.03\% & 3.51 & 13.51 \\ 
Societe Generale (GLE) & 0.97\% & 1.19\% & 6.57\% & 0.03\% & 1.75 & 2.44 \\ 
St. Galler Kantonalbank (SGKN)& 0.07\% & 0.06\% & 0.28\% & 0.01\% & 1.52 & 1.3 \\ 
Standard Chartered Plc (STAN) & 0.41\% & 0.64\% & 3.23\% & 0.03\% & 2.45 & 5.21 \\ 
Swedbank (SWED) & 0.92\% & 2.4\% & 22.87\% & 0.03\% & 4.49 & 23.1 \\ 
Unicredit (UCG) & 1.11\% & 1.43\% & 8.36\% & 0.02\% & 1.63 & 1.96 \\ 
\hline
\end{tabular}
}
\end{center}
\end{table}

\begin{table}[H]
\caption{Summary of Default Probabilities by Country.}\label{PDbyCountry}
\centering
\scalebox{0.9}{
\begin{tabular}{lcccccc}
\hline
& Mean & SD & Maximum & Minimum & Skewness & Kurtosis \\ 
\hline
Italy & 1.06\% & 1.12\% & 5.71\% & 0.03\% & 1.48 & 2.06 \\ 
UK & 0.91\% & 1.37\% & 9.82\% & 0.03\% & 2.89 & 9.56 \\ 
Spain & 0.47\% & 0.48\% & 2.58\% & 0.01\% & 1.41 & 1.55 \\ 
Switzerland & 0.22\% & 0.22\% & 1.18\% & 0.03\% & 1.81 & 3.21 \\ 
France & 0.88\% & 1.08\% & 5.4\% & 0.03\% & 1.78 & 2.37 \\ 
Germany & 1.51\% & 2.21\% & 15.95\% & 0.07\% & 3.1 & 11.56 \\ 
Denmark & 0.78\% & 1.71\% & 16.21\% & 0.01\% & 4.47 & 24.12 \\ 
Norway & 0.54\% & 1.13\% & 9.11\% & 0.03\% & 3.93 & 15.93 \\ 
Austria & 0.93\% & 1.39\% & 9.4\% & 0.06\% & 2.73 & 8.07 \\ 
Belgium & 1.17\% & 2.09\% & 14.73\% & 0.02\% & 2.81 & 8.27 \\ 
Sweden & 0.58\% & 1.28\% & 11.07\% & 0.03\% & 4.1 & 18.95 \\ 
Netherlands & 1.12\% & 2.15\% & 21.55\% & 0.04\% & 3.99 & 20.3 \\ 
\hline
\end{tabular}
}
\end{table}

\newpage
\bibliographystyle{elsarticle-harv}
\bibliography{referencesnew}

\begin{thebibliography}{43}
\expandafter\ifx\csname natexlab\endcsname\relax\def\natexlab#1{#1}\fi
\providecommand{\url}[1]{\texttt{#1}}
\providecommand{\href}[2]{#2}
\providecommand{\path}[1]{#1}
\providecommand{\DOIprefix}{doi:}
\providecommand{\ArXivprefix}{arXiv:}
\providecommand{\URLprefix}{URL: }
\providecommand{\Pubmedprefix}{pmid:}
\providecommand{\doi}[1]{\href{http://dx.doi.org/#1}{\path{#1}}}
\providecommand{\Pubmed}[1]{\href{pmid:#1}{\path{#1}}}
\providecommand{\bibinfo}[2]{#2}
\ifx\xfnm\relax \def\xfnm[#1]{\unskip,\space#1}\fi
\bibitem[{Acemoglu et~al.(2015)Acemoglu, Ozdaglar and
  Tahbaz-Salehi}]{acemoglu2015systemic}
\bibinfo{author}{Acemoglu, D.}, \bibinfo{author}{Ozdaglar, A.},
  \bibinfo{author}{Tahbaz-Salehi, A.}, \bibinfo{year}{2015}.
\newblock \bibinfo{title}{Systemic risk and stability in financial networks}.
\newblock \bibinfo{journal}{American Economic Review} \bibinfo{volume}{105},
  \bibinfo{pages}{564--608}.
\bibitem[{Acharya et~al.(2014)Acharya, Drechsler and
  Schnabl}]{acharya2014pyrrhic}
\bibinfo{author}{Acharya, V.}, \bibinfo{author}{Drechsler, I.},
  \bibinfo{author}{Schnabl, P.}, \bibinfo{year}{2014}.
\newblock \bibinfo{title}{A pyrrhic victory? bank bailouts and sovereign credit
  risk}.
\newblock \bibinfo{journal}{The Journal of Finance} \bibinfo{volume}{69},
  \bibinfo{pages}{2689--2739}.
\bibitem[{Acharya et~al.(2012)Acharya, Engle and
  Richardson}]{acharya2012capital}
\bibinfo{author}{Acharya, V.}, \bibinfo{author}{Engle, R.},
  \bibinfo{author}{Richardson, M.}, \bibinfo{year}{2012}.
\newblock \bibinfo{title}{Capital shortfall: A new approach to ranking and
  regulating systemic risks}.
\newblock \bibinfo{journal}{American Economic Review} \bibinfo{volume}{102},
  \bibinfo{pages}{59--64}.
\bibitem[{Aiyar(2012)}]{aiyar2012financial}
\bibinfo{author}{Aiyar, S.}, \bibinfo{year}{2012}.
\newblock \bibinfo{title}{From financial crisis to great recession: The role of
  globalized banks}.
\newblock \bibinfo{journal}{American Economic Review} \bibinfo{volume}{102},
  \bibinfo{pages}{225--30}.
\bibitem[{Andrieu et~al.(2003)Andrieu, De~Freitas, Doucet and
  Jordan}]{andrieu2003introduction}
\bibinfo{author}{Andrieu, C.}, \bibinfo{author}{De~Freitas, N.},
  \bibinfo{author}{Doucet, A.}, \bibinfo{author}{Jordan, M.I.},
  \bibinfo{year}{2003}.
\newblock \bibinfo{title}{An introduction to {MCMC} for machine learning}.
\newblock \bibinfo{journal}{Machine learning} \bibinfo{volume}{50},
  \bibinfo{pages}{5--43}.
\bibitem[{Bartram et~al.(2007)Bartram, Brown and Hund}]{bartram2007estimating}
\bibinfo{author}{Bartram, S.M.}, \bibinfo{author}{Brown, G.W.},
  \bibinfo{author}{Hund, J.E.}, \bibinfo{year}{2007}.
\newblock \bibinfo{title}{Estimating systemic risk in the international
  financial system}.
\newblock \bibinfo{journal}{Journal of Financial Economics}
  \bibinfo{volume}{86}, \bibinfo{pages}{835--869}.
\bibitem[{Battiston et~al.(2012)Battiston, Puliga, Kaushik, Tasca and
  Caldarelli}]{battiston2012debtrank}
\bibinfo{author}{Battiston, S.}, \bibinfo{author}{Puliga, M.},
  \bibinfo{author}{Kaushik, R.}, \bibinfo{author}{Tasca, P.},
  \bibinfo{author}{Caldarelli, G.}, \bibinfo{year}{2012}.
\newblock \bibinfo{title}{Debtrank: Too central to fail? financial networks,
  the fed and systemic risk}.
\newblock \bibinfo{journal}{Scientific reports} \bibinfo{volume}{2},
  \bibinfo{pages}{541}.
\bibitem[{Billio et~al.(2018)Billio, Caporin, Frattarolo and
  Pelizzon}]{billio2018networks}
\bibinfo{author}{Billio, M.}, \bibinfo{author}{Caporin, M.},
  \bibinfo{author}{Frattarolo, L.}, \bibinfo{author}{Pelizzon, L.},
  \bibinfo{year}{2018}.
\newblock \bibinfo{title}{Networks in risk spillovers: A multivariate garch
  perspective} .
\bibitem[{Birch et~al.(2016)Birch, Pantelous and
  Soram{\"a}ki}]{birch2016analysis}
\bibinfo{author}{Birch, J.}, \bibinfo{author}{Pantelous, A.A.},
  \bibinfo{author}{Soram{\"a}ki, K.}, \bibinfo{year}{2016}.
\newblock \bibinfo{title}{Analysis of correlation based networks representing
  dax 30 stock price returns}.
\newblock \bibinfo{journal}{Computational Economics} \bibinfo{volume}{47},
  \bibinfo{pages}{501--525}.
\bibitem[{Brida and Risso(2010)}]{brida2010dynamics}
\bibinfo{author}{Brida, J.G.}, \bibinfo{author}{Risso, W.A.},
  \bibinfo{year}{2010}.
\newblock \bibinfo{title}{Dynamics and structure of the 30 largest north
  american companies}.
\newblock \bibinfo{journal}{Computational Economics} \bibinfo{volume}{35},
  \bibinfo{pages}{85}.
\bibitem[{Chi et~al.(2010)Chi, Liu and Lau}]{chi2010network}
\bibinfo{author}{Chi, K.T.}, \bibinfo{author}{Liu, J.}, \bibinfo{author}{Lau,
  F.C.}, \bibinfo{year}{2010}.
\newblock \bibinfo{title}{A network perspective of the stock market}.
\newblock \bibinfo{journal}{Journal of Empirical Finance} \bibinfo{volume}{17},
  \bibinfo{pages}{659--667}.
\bibitem[{Das(2016)}]{das2016matrix}
\bibinfo{author}{Das, S.R.}, \bibinfo{year}{2016}.
\newblock \bibinfo{title}{Matrix metrics: Network-based systemic risk scoring}.
\newblock \bibinfo{journal}{The Journal of Alternative Investments}
  \bibinfo{volume}{18}, \bibinfo{pages}{33--51}.
\bibitem[{De~Haas and Van~Horen(2012)}]{de2012international}
\bibinfo{author}{De~Haas, R.}, \bibinfo{author}{Van~Horen, N.},
  \bibinfo{year}{2012}.
\newblock \bibinfo{title}{International shock transmission after the {Lehman}
  brothers collapse: Evidence from syndicated lending}.
\newblock \bibinfo{journal}{American Economic Review} \bibinfo{volume}{102},
  \bibinfo{pages}{231--37}.
\bibitem[{Durante et~al.(2017)Durante, Dunson and
  Vogelstein}]{durante2017nonparametric}
\bibinfo{author}{Durante, D.}, \bibinfo{author}{Dunson, D.B.},
  \bibinfo{author}{Vogelstein, J.T.}, \bibinfo{year}{2017}.
\newblock \bibinfo{title}{Nonparametric bayes modeling of populations of
  networks}.
\newblock \bibinfo{journal}{Journal of the American Statistical Association}
  \bibinfo{volume}{112}, \bibinfo{pages}{1516--1530}.
\bibitem[{Elliott et~al.(2014)Elliott, Golub and
  Jackson}]{elliott2014financial}
\bibinfo{author}{Elliott, M.}, \bibinfo{author}{Golub, B.},
  \bibinfo{author}{Jackson, M.O.}, \bibinfo{year}{2014}.
\newblock \bibinfo{title}{Financial networks and contagion}.
\newblock \bibinfo{journal}{American Economic Review} \bibinfo{volume}{104},
  \bibinfo{pages}{3115--53}.
\bibitem[{Engle et~al.(2014)Engle, Jondeau and Rockinger}]{engle2014systemic}
\bibinfo{author}{Engle, R.}, \bibinfo{author}{Jondeau, E.},
  \bibinfo{author}{Rockinger, M.}, \bibinfo{year}{2014}.
\newblock \bibinfo{title}{Systemic risk in {Europe}}.
\newblock \bibinfo{journal}{Review of Finance} \bibinfo{volume}{19},
  \bibinfo{pages}{145--190}.
\bibitem[{Flavin et~al.(2002)Flavin, Hurley and
  Rousseau}]{flavin2002explaining}
\bibinfo{author}{Flavin, T.J.}, \bibinfo{author}{Hurley, M.J.},
  \bibinfo{author}{Rousseau, F.}, \bibinfo{year}{2002}.
\newblock \bibinfo{title}{Explaining stock market correlation: A gravity model
  approach}.
\newblock \bibinfo{journal}{The Manchester School} \bibinfo{volume}{70},
  \bibinfo{pages}{87--106}.
\bibitem[{Friel et~al.(2016)Friel, Rastelli, Wyse and
  Raftery}]{friel2016interlocking}
\bibinfo{author}{Friel, N.}, \bibinfo{author}{Rastelli, R.},
  \bibinfo{author}{Wyse, J.}, \bibinfo{author}{Raftery, A.E.},
  \bibinfo{year}{2016}.
\newblock \bibinfo{title}{Interlocking directorates in {Irish} companies using
  a latent space model for bipartite networks}.
\newblock \bibinfo{journal}{Proceedings of the National Academy of Sciences}
  \bibinfo{volume}{113}, \bibinfo{pages}{6629--6634}.
\bibitem[{Friel and Wyse(2012)}]{friel2012estimating}
\bibinfo{author}{Friel, N.}, \bibinfo{author}{Wyse, J.}, \bibinfo{year}{2012}.
\newblock \bibinfo{title}{Estimating the evidence--a review}.
\newblock \bibinfo{journal}{Statistica Neerlandica} \bibinfo{volume}{66},
  \bibinfo{pages}{288--308}.
\bibitem[{Giner et~al.(2018)Giner, Mendoza~Aguilar and
  Morini-Marrero}]{giner2018correlation}
\bibinfo{author}{Giner, J.}, \bibinfo{author}{Mendoza~Aguilar, J.},
  \bibinfo{author}{Morini-Marrero, S.}, \bibinfo{year}{2018}.
\newblock \bibinfo{title}{Correlation as probability: applications of
  {Sheppard’s formula} to financial assets}.
\newblock \bibinfo{journal}{Quantitative Finance} \bibinfo{volume}{18},
  \bibinfo{pages}{777--787}.
\bibitem[{Giudici and Parisi(2016)}]{Paolo2016}
\bibinfo{author}{Giudici, P.}, \bibinfo{author}{Parisi, L.},
  \bibinfo{year}{2016}.
\newblock \bibinfo{title}{CoRisk: measuring systemic risk through default
  probability contagion}.
\newblock \bibinfo{type}{DEM Working Papers Series} \bibinfo{number}{116}.
  University of Pavia, Department of Economics and Management.
\newblock \URLprefix
  \url{https://EconPapers.repec.org/RePEc:pav:demwpp:demwp0116}.
\bibitem[{Grassi et~al.(2010)Grassi, Stefani and
  Torriero}]{grassi2010centrality}
\bibinfo{author}{Grassi, R.}, \bibinfo{author}{Stefani, S.},
  \bibinfo{author}{Torriero, A.}, \bibinfo{year}{2010}.
\newblock \bibinfo{title}{Centrality in organizational networks}.
\newblock \bibinfo{journal}{International journal of intelligent systems}
  \bibinfo{volume}{25}, \bibinfo{pages}{253--265}.
\bibitem[{Handcock et~al.(2007)Handcock, Raftery and
  Tantrum}]{handcock2007model}
\bibinfo{author}{Handcock, M.S.}, \bibinfo{author}{Raftery, A.E.},
  \bibinfo{author}{Tantrum, J.M.}, \bibinfo{year}{2007}.
\newblock \bibinfo{title}{Model-based clustering for social networks}.
\newblock \bibinfo{journal}{Journal of the Royal Statistical Society: Series A
  (Statistics in Society)} \bibinfo{volume}{170}, \bibinfo{pages}{301--354}.
\bibitem[{Hledik and Rastelli(2018)}]{hledik2018dynamic}
\bibinfo{author}{Hledik, J.}, \bibinfo{author}{Rastelli, R.},
  \bibinfo{year}{2018}.
\newblock \bibinfo{title}{A dynamic network model to measure exposure
  diversification in the {Austrian interbank} market}.
\newblock \bibinfo{journal}{arXiv preprint arXiv:1804.01367} .
\bibitem[{Hoff et~al.(2002)Hoff, Raftery and Handcock}]{hoff2002latent}
\bibinfo{author}{Hoff, P.D.}, \bibinfo{author}{Raftery, A.E.},
  \bibinfo{author}{Handcock, M.S.}, \bibinfo{year}{2002}.
\newblock \bibinfo{title}{Latent space approaches to social network analysis}.
\newblock \bibinfo{journal}{Journal of the American Statistical Association}
  \bibinfo{volume}{97}, \bibinfo{pages}{1090--1098}.
\bibitem[{Kenett et~al.(2015)Kenett, Huang, Vodenska, Havlin and
  Stanley}]{kenett2015partial}
\bibinfo{author}{Kenett, D.Y.}, \bibinfo{author}{Huang, X.},
  \bibinfo{author}{Vodenska, I.}, \bibinfo{author}{Havlin, S.},
  \bibinfo{author}{Stanley, H.E.}, \bibinfo{year}{2015}.
\newblock \bibinfo{title}{Partial correlation analysis: Applications for
  financial markets}.
\newblock \bibinfo{journal}{Quantitative Finance} \bibinfo{volume}{15},
  \bibinfo{pages}{569--578}.
\bibitem[{Mantegna and Stanley(1999)}]{mantegna1999introduction}
\bibinfo{author}{Mantegna, R.N.}, \bibinfo{author}{Stanley, H.E.},
  \bibinfo{year}{1999}.
\newblock \bibinfo{title}{Introduction to econophysics: correlations and
  complexity in finance}.
\newblock \bibinfo{publisher}{Cambridge university press}.
\bibitem[{Matesanz and Ortega(2014)}]{matesanz2014network}
\bibinfo{author}{Matesanz, D.}, \bibinfo{author}{Ortega, G.J.},
  \bibinfo{year}{2014}.
\newblock \bibinfo{title}{Network analysis of exchange data: Interdependence
  drives crisis contagion}.
\newblock \bibinfo{journal}{Quality \& Quantity} \bibinfo{volume}{48},
  \bibinfo{pages}{1835--1851}.
\bibitem[{M{\o}ller et~al.(2006)M{\o}ller, Pettitt, Reeves and
  Berthelsen}]{moller2006efficient}
\bibinfo{author}{M{\o}ller, J.}, \bibinfo{author}{Pettitt, A.N.},
  \bibinfo{author}{Reeves, R.}, \bibinfo{author}{Berthelsen, K.K.},
  \bibinfo{year}{2006}.
\newblock \bibinfo{title}{An efficient {Markov chain Monte Carlo} method for
  distributions with intractable normalising constants}.
\newblock \bibinfo{journal}{Biometrika} \bibinfo{volume}{93},
  \bibinfo{pages}{451--458}.
\bibitem[{Murray et~al.(2012)Murray, Ghahramani and MacKay}]{murray2012mcmc}
\bibinfo{author}{Murray, I.}, \bibinfo{author}{Ghahramani, Z.},
  \bibinfo{author}{MacKay, D.}, \bibinfo{year}{2012}.
\newblock \bibinfo{title}{Mcmc for doubly-intractable distributions}.
\newblock \bibinfo{journal}{arXiv preprint arXiv:1206.6848} .
\bibitem[{Pourkhanali et~al.(2016)Pourkhanali, Kim, Tafakori and
  Fard}]{pourkhanali2016measuring}
\bibinfo{author}{Pourkhanali, A.}, \bibinfo{author}{Kim, J.M.},
  \bibinfo{author}{Tafakori, L.}, \bibinfo{author}{Fard, F.A.},
  \bibinfo{year}{2016}.
\newblock \bibinfo{title}{Measuring systemic risk using vine-copula}.
\newblock \bibinfo{journal}{Economic modelling} \bibinfo{volume}{53},
  \bibinfo{pages}{63--74}.
\bibitem[{Qi~X.(2013)}]{Qi2013}
\bibinfo{author}{Qi~X., Duval~R.D., C.K.F.E.S.A.W.Q.W.Y.T.W.Z.C.},
  \bibinfo{year}{2013}.
\newblock \bibinfo{title}{Terrorist networks, network energy and node removal:
  a new measure of centrality based on laplacian energy} .
\bibitem[{Raftery et~al.(2012)Raftery, Niu, Hoff and Yeung}]{raftery2012fast}
\bibinfo{author}{Raftery, A.E.}, \bibinfo{author}{Niu, X.},
  \bibinfo{author}{Hoff, P.D.}, \bibinfo{author}{Yeung, K.Y.},
  \bibinfo{year}{2012}.
\newblock \bibinfo{title}{Fast inference for the latent space network model
  using a case-control approximate likelihood}.
\newblock \bibinfo{journal}{Journal of Computational and Graphical Statistics}
  \bibinfo{volume}{21}, \bibinfo{pages}{901--919}.
\bibitem[{Rastelli et~al.(2016)Rastelli, Friel and
  Raftery}]{rastelli2016properties}
\bibinfo{author}{Rastelli, R.}, \bibinfo{author}{Friel, N.},
  \bibinfo{author}{Raftery, A.E.}, \bibinfo{year}{2016}.
\newblock \bibinfo{title}{Properties of latent variable network models}.
\newblock \bibinfo{journal}{Network Science} \bibinfo{volume}{4},
  \bibinfo{pages}{407--432}.
\bibitem[{Rastelli et~al.(2018)Rastelli, Maire and
  Friel}]{rastelli2018computationally}
\bibinfo{author}{Rastelli, R.}, \bibinfo{author}{Maire, F.},
  \bibinfo{author}{Friel, N.}, \bibinfo{year}{2018}.
\newblock \bibinfo{title}{Computationally efficient inference for latent
  position network models}.
\newblock \bibinfo{journal}{arXiv preprint arXiv:1804.02274} .
\bibitem[{Scardoni and Laudanna(2012)}]{scardoni2012centralities}
\bibinfo{author}{Scardoni, G.}, \bibinfo{author}{Laudanna, C.},
  \bibinfo{year}{2012}.
\newblock \bibinfo{title}{Centralities based analysis of complex networks}.
\newblock \bibinfo{journal}{New Frontiers in Graph Theory} ,
  \bibinfo{pages}{323--348}.
\bibitem[{Schweitzer et~al.(2009)Schweitzer, Fagiolo, Sornette, Vega-Redondo,
  Vespignani and White}]{schweitzer2009economic}
\bibinfo{author}{Schweitzer, F.}, \bibinfo{author}{Fagiolo, G.},
  \bibinfo{author}{Sornette, D.}, \bibinfo{author}{Vega-Redondo, F.},
  \bibinfo{author}{Vespignani, A.}, \bibinfo{author}{White, D.R.},
  \bibinfo{year}{2009}.
\newblock \bibinfo{title}{Economic networks: The new challenges}.
\newblock \bibinfo{journal}{science} \bibinfo{volume}{325},
  \bibinfo{pages}{422--425}.
\bibitem[{Shortreed et~al.(2006)Shortreed, Handcock and
  Hoff}]{shortreed2006positional}
\bibinfo{author}{Shortreed, S.}, \bibinfo{author}{Handcock, M.S.},
  \bibinfo{author}{Hoff, P.}, \bibinfo{year}{2006}.
\newblock \bibinfo{title}{Positional estimation within a latent space model for
  networks}.
\newblock \bibinfo{journal}{Methodology} \bibinfo{volume}{2},
  \bibinfo{pages}{24--33}.
\bibitem[{Wang and Xie(2015)}]{wang2015correlation}
\bibinfo{author}{Wang, G.J.}, \bibinfo{author}{Xie, C.}, \bibinfo{year}{2015}.
\newblock \bibinfo{title}{Correlation structure and dynamics of international
  real estate securities markets: A network perspective}.
\newblock \bibinfo{journal}{Physica A: Statistical Mechanics and its
  Applications} \bibinfo{volume}{424}, \bibinfo{pages}{176--193}.
\bibitem[{Wang and Xie(2016)}]{wang2016tail}
\bibinfo{author}{Wang, G.J.}, \bibinfo{author}{Xie, C.}, \bibinfo{year}{2016}.
\newblock \bibinfo{title}{Tail dependence structure of the foreign exchange
  market: A network view}.
\newblock \bibinfo{journal}{Expert Systems with Applications}
  \bibinfo{volume}{46}, \bibinfo{pages}{164--179}.
\bibitem[{Wang et~al.(2018)Wang, Xie and Stanley}]{wang2018correlation}
\bibinfo{author}{Wang, G.J.}, \bibinfo{author}{Xie, C.},
  \bibinfo{author}{Stanley, H.E.}, \bibinfo{year}{2018}.
\newblock \bibinfo{title}{Correlation structure and evolution of world stock
  markets: Evidence from pearson and partial correlation-based networks}.
\newblock \bibinfo{journal}{Computational Economics} \bibinfo{volume}{51},
  \bibinfo{pages}{607--635}.
\bibitem[{Wen et~al.(2019)Wen, Yang and Zhou}]{wen2019tail}
\bibinfo{author}{Wen, F.}, \bibinfo{author}{Yang, X.}, \bibinfo{author}{Zhou,
  W.X.}, \bibinfo{year}{2019}.
\newblock \bibinfo{title}{Tail dependence networks of global stock markets}.
\newblock \bibinfo{journal}{International Journal of Finance \& Economics}
  \bibinfo{volume}{24}, \bibinfo{pages}{558--567}.
\bibitem[{Winkler(2012)}]{winkler2012image}
\bibinfo{author}{Winkler, G.}, \bibinfo{year}{2012}.
\newblock \bibinfo{title}{Image analysis, random fields and Markov chain Monte
  Carlo methods: a mathematical introduction}. volume~\bibinfo{volume}{27}.
\newblock \bibinfo{publisher}{Springer Science \& Business Media}.

\end{thebibliography}
\end{document}